# Modeling Interactome: Scale-Free or Geometric?

Nataša Pržulj,[1] Derek G. Corneil,[1] Igor Jurisica[2,1,*]

[1]Department of Computer Science, University of Toronto,
10 King's College Road, Toronto, ON, M5S 3G4, Canada
[2]Ontario Cancer Institute, University Health Network, Division of Cancer Informatics
610 University Avenue, Toronto, ON, M5G 2M9, Canada

[*]To whom correspondence should be addressed; E-mail: juris@cs.toronto.edu.

**Networks have been used to model many real-world phenomena to better understand the phenomena and to guide experiments in order to predict their behavior. Since incorrect models lead to incorrect predictions, it is vital to have a correct model. As a result, new techniques and models for analyzing and modeling real-world networks have recently been introduced. One example of large and complex networks involves protein-protein interaction (PPI) networks. We demonstrate that the currently popular scale-free model of PPI networks fails to fit the data in several respects. We show that a random geometric model provides a much more accurate model of the PPI data.**

Many real-world phenomena have been modeled by large *networks* including the World Wide Web, electronic circuits, collaborations between scientists, metabolic pathways, and protein-protein interactions (PPIs). A common property of these phenomena is that they all consist of components (modeled by network *nodes*) and pairwise interactions between the components (modeled by links between the nodes, i.e., by network *edges*). Studying statistical and theo-retical properties of large networks (also called *graphs*) has gained considerable attention in the past few years. Various network models have been proposed to describe properties of large real-world networks, starting with the earliest models of Erdös-Rényi random graphs[1, 2, 3] and including more recent small-world[4], scale-free[5], and hierarchical[6] models. Excellent review papers have recently appeared describing this emerging, large research area[7, 8, 9, 10].

Despite recent significant advances in understanding large real-world networks, this area of research is still in its infancy[8, 7]. Novel techniques for analyzing, characterizing, and modeling structures of these networks need to be developed. As new data becomes available, we must ensure that the theoretical models are still accurately representing the data. The scale-free model has been assumed to accurately model PPI networks. Current scale-free model of human PPI network has been used for planning experiments in order to optimize time and cost required for their completion[11]. In particular, the model was used to form the basis of an algorithmic strategy for guiding experiments which would detect up to $90\%$ of the human interactome with less than a third of the proteome used as bait in high-throughput pull-down experiments[11]. Therefore, having an improved model for PPI networks will have significant biological implications.

This paper uses a method for detecting local structural properties of large networks and proposes a new model of PPI networks. Our new measure of local network structure consists of 29 network measurements. We find that, using this new measure of network structure, the PPI networks of *S. cerevisiae* and *D. melanogaster* are more accurately modeled by geometric random graphs than by the scale-free model. The extent of this improvement is such that even perturbing the network by random additions, deletions and rewiring of $30\%$ of the edges introduces much smaller error when compared to the error from modeling the network by scale-free, or other currently available network models (details are provided below). In addition, we show that three out of four standard network parameters measuring a global network structure also show an improved fit between the experimentally-determined PPI networks and the geometric random graph model than between the PPI networks and the scale-free model.

The most commonly studied statistical properties of large networks, measuring their global structure, are the degree distribution, network diameter, and clustering coefficients, defined as follows. The *degree* of a node is the number of edges (connections) incident to the node. The *degree distribution*, $P(k)$, describes the probability that a node has degree $k$. This network property has been used to distinguish among different network models; in particular, Erdös-Rényi random networks have a Poisson degree distribution, while *scale-free* networks have a power-law degree distribution $P(k) \sim k^{-\gamma}$, where $\gamma$ is a positive number. The smallest number of links that have to be traversed to get from node $x$ to node $y$ in a network is called the *distance* between nodes $x$ and $y$ and a path through the network that achieves this distance is called a *shortest path* between $x$ and $y$. The average of shortest path lengths over all pairs of nodes in a network is called the network *diameter*. (Note that in classical graph theory, the diameter is the maximum of shortest path lengths over all pairs of nodes in the network[12].) This network property also distinguishes different network models: for example, the diameter of Erdös-Rényi random networks on $n$ nodes is proportional to $\log n$, the network property often referred to as the *small-world* property; the diameters of scale-free random networks with degree exponent $2 < \gamma < 3$, which have been observed for most real-world networks, are *ultra-small*[13, 14], i.e., proportional to $\log \log n$. The *clustering coefficient of node $v$* in a network is defined as $C_v = \frac{2e_1}{n_1(n_1-1)}$, where $v$ is linked to $n_1$ neighboring nodes and $e_1$ is the number of edges amongst the $n_1$ neighbors of $v$. The average of $C_v$ over all nodes $v$ of a network is the *clustering coefficient $C$* of the whole network and it measures the tendency of the network to form highly interconnected regions called clusters. The average clustering coefficient of all nodes of degree $k$ in a network, $C(k)$, has been shown to follow $C(k) \sim k^{-1}$ for many real-world networks indicating a network's hierarchical structure[15, 6]. Many real world networks have been shown to have high clustering coefficients and to exhibit small-world and scale-free properties.

In addition to the above global properties of network structure, a new bottom-up approach focusing on finding small, over-represented patterns in a network has recently been introduced[16, 17, 18, 19]. In this approach, all small *subgraphs* (subnetworks whose nodes and edges belong to the large network) of a large network are identified and the ones that appear in the network significantly more frequently than in the randomized network are called network *motifs*. Different types of real-world networks have been shown to have different motifs[16]. Furthermore, different real-world evolved and designed networks have been grouped into superfamilies according to their local structural properties[19].

Recently, there has been a lot of interest in the properties of PPI networks. PPI networks for the yeast *S. cerevisiae* resulting from different high-throughput studies[20, 21, 22] have been shown to have scale-free degree distributions[23, 24]. They have hierarchical structures with $C(k)$ scaling as $k^{-1}$ [25]. The *S. cerevisiae* PPI network constructed on combined, mostly two-hybrid analysis data[20, 21], has been shown to have two network motifs[16], those corresponding to graphs 2 and 4 presented in Figure 1. The degree distributions of this yeast PPI network, as well as the PPI network of the bacterium *Helicobacter pylori*, have been shown to decay according to a power law[23, 26]. However, the high-confidence and a larger fruitfly *D. melanogaster* PPI networks have been shown to decay close to, but faster than a power law[27]. Furthermore, the shortest path distribution and the frequencies of 3-15-node cycles in the high-confidence fruitfly PPI network have been shown to differ from those of randomly rewired networks which preserve the same degree distribution as the original PPI network[27]. When studying PPI networks, it should be noted that all of the current publicly available PPI data sets contain a percentage of false positives and are also largely incomplete, i.e, the number of false negatives is arguably much larger than the number of false positives. Since the genomes of many species have already been sequenced, it is expected that the predicted number of proteins in PPI data sets will not change significantly, but the number of known interactions will grow dramatically.

Our approach to analyzing large networks belongs to the bottom-up type. Similar to the approach of Milo *et al.*[19], we identify all 3-5-node subgraphs of PPI networks for *S. cerevisiae* and *D. melanogaster*. We compare the frequencies of the appearance of these subgraphs in PPI networks with the frequencies of their appearance in four different types of random networks: (a) Erdös-Rényi random networks with the same number of nodes and edges as the corresponding PPI networks (ER); (b) Erdös-Rényi random networks with the same number of nodes, edges, and the same degree distribution as corresponding PPI networks (ER-DD); (c) scale-free random networks with the same number of nodes and the number of edges within $1\%$ of those of the corresponding PPI networks (SF); and (d) several types of geometric random graphs with the number of nodes and the number of edges within $1\%$ of those of the corresponding PPI networks (GEO) (see Supplementary Information). In the *geometric random graph* model, nodes correspond to independently and uniformly randomly distributed points in a metric space, and two nodes are linked by an edge if the distance between them is smaller than or equal to some radius $r$, where distance is an arbitrary distance norm in the metric space (more details about geometric random graphs can be found in[28]). We used three different geometric random graph models, defining points uniformly at random in 2-dimensional Euclidean space (GEO-2D), 3-dimensional Euclidean space (GEO-3D), and 4-dimensional Euclidean space (GEO-4D); the Euclidean distance measure between the points was used to determine if two points are close enough to be linked by an edge in the corresponding geometric random graph (see Supplementary Information). To our knowledge, this study is the first one to use geometric random graphs to model PPI networks.

The number of different connected networks on $n$ nodes increases exponentially with $n$. For $n = 3, 4$, and $5$, there are $2, 6$, and $21$ different connected networks on $n$ nodes respectively. To avoid terminology confusing network motifs with network subgraphs (motifs are special types of subgraphs), we use the term *graphlet* to denote a connected network with a small number of nodes. All 3-5-node graphlets are presented in Figure 1. (Note that in their analysis of undirected networks, Milo *et al.*[19] examined the presence of the 8 graphlets of size 3 or 4.) We use the *graphlet frequency*, i.e., the number of occurrences of a graphlet in a network, as a new network parameter and show that PPI networks are closest to geometric random graphs with respect to this new network parameter (details are given below). In addition, despite the difference in degree distributions of PPI networks and geometric random graphs and the similarity between degree distributions of PPI networks and scale-free networks, we show that the diameter and clustering coefficient parameters also indicate that PPI networks are closer to the geometric random graph model than to the ER, ER-DD and SF models. We hypothesize that the discrepancy between the degree distributions of PPI and GEO networks is caused by a high percentage of false negatives in the PPI networks and that when PPI data sets become denser and more complete, the degree distributions of PPI networks will be closer to Poisson distributions, characteristic of geometric random graphs.

We analyzed graphlet frequencies of four PPI networks: the high-confidence yeast *S. cerevisiae* PPI network involving 2455 interactions amongst 988 proteins[29]; the yeast *S. cerevisiae* PPI network involving 11000 interactions amongst 2401 proteins[29] (these are the top 11000 interactions in von Mering *et al.* classification[29]); the high-confidence fruitfly *D. melanogaster* PPI network involving 4637 interactions amongst 4602 proteins[27]; and the entire fruitfly *D. melanogaster* PPI network as published in[27] involving 20007 interactions amongst 6985 proteins which includes low confidence interactions. We computed graphlet frequencies in the PPI and the corresponding random networks of the previously described four different types. Graphlet counts quantify the local structural properties of a network. Currently, our knowledge of the connections in PPI networks is incomplete (i.e., we do not know all the edges). The edges we *do* know are dominated by lab studies of connectivity in the vicinity of proteins that are currently considered "important". However, we hypothesize that the local structural properties of the full PPI network, once all connections are made, are similar to the local structural properties of the currently known, highly studied parts of the network. Thus, we would expect that the *relative* frequency of graphlets among the currently known connections is similar to the relative frequency of graphlets in the full PPI network, which is as yet unknown. Thus, we use the *relative frequency of graphlets* $N_i(G)/T(G)$ to characterize PPI networks and the networks we use to model them, where $N_i(G)$ is the number of graphlets of type $i$ ($i \in \{1, \ldots, 29\}$) in a network $G$, and $T(G) = \sum_{i=1}^{29} N_i(G)$ is the total number of graphlets of $G$. In this model, then, the "similarity" between two graphs should be independent of the total number of edges, and should depend only upon the differences between relative frequencies of graphlets. Thus, we define the *relative graphlet frequency distance* $D(G, H)$, or *distance* for brevity, between two graphs $G$ and $H$ as

$$D(G, H) = \sum_{i=1}^{29} |F_i(G) - F_i(H)|,$$

where $F_i(G) = -\log(N_i(G)/T(G))$. We use the logarithm of the graphlet frequency because frequencies of different graphlets can differ by several orders of magnitude and we do not want the distance measure to be entirely dominated by the most frequent graphlets.

Using this method, we computed the distances between several real-world PPI networks and the corresponding ER, ER-DD, SF, and GEO random networks. We found that the GEO random networks fit the data an order of magnitude better in the higher-confidence PPI networks, and less so (but still better) in the more noisy PPI networks (see Supplementary Table 2 of the Supplementary Information). The only exception is the larger fruitfly PPI network, with about 77% of its edges corresponding to lower confidence interactions[27]; this PPI network is about 2.7 times closer to the scale-free than to the geometric network model with respect to this parameter (see Supplementary Information). We hypothesize that this behavior of the graphlet frequency parameter is the consequence of a large amount of noise present in this fruitfly PPI network;

our analysis of the diameters and clustering coefficients of these networks further support this hypothesis (see below).

An illustration showing graphlet frequencies in the high-confidence yeast PPI network and the corresponding random model networks is presented in Figure 2. To obtain a geometric random network model that closely fits the graphlet frequency parameter of this PPI network (see Fig. 2 F and Supplementary Information), we constructed two sets of 3-dimensional geometric random networks with the same number of nodes, but about three and six times as many edges as the PPI network, respectively. We did this because, as mentioned above, the current yeast high-confidence PPI network is missing many edges, so we expect that the complete PPI networks would be much denser. Also, by making the GEO-3D networks corresponding to this PPI network about six times as dense as the PPI network, we matched the maximum degree of the PPI network to those of these geometric random networks. We believe that the maximum degree of this PPI network is not likely to change significantly due to the extent of research being done on the highly connected regions of the network.

Since PPI networks contain noise, we examined the robustness of the graphlet frequency parameter by adding noise to the yeast high-confidence PPI network and comparing graphlet frequencies of the perturbed networks and the PPI network. In particular, we perturbed this PPI network by randomly adding, deleting, and rewiring 10, 20, and 30 percent of its edges. We computed distances between the perturbed networks and the PPI network by using the distance function defined above. We found the exceptional robustness of the graphlet frequency parameter to random additions of edges very encouraging, especially in light of the currently available PPI networks missing many edges. In particular, additions of 30% of edges resulted in networks which were about 21 times closer to the PPI network than the corresponding SF random networks. We also found that graphlet frequencies were fairly robust to random edge deletions and rewirings (deletions and rewirings of 30% of edges resulted in networks which were about 6 times closer to the PPI network than the corresponding SF random networks), which further increases our confidence in PPI networks having geometric properties despite the presence of false positives in the data (see Supplementary Information).

When we compared the commonly studied statistical properties of large networks, namely the degree distribution, network diameter, and clustering coefficients $C$ and $C(k)$, of the PPI and the corresponding model networks, we found that all of the parameters of the two yeast PPI networks, except degree distributions, matched those of geometric random graphs. An illustration of the behavior of $C(k)$ in the PPI and the model networks is presented in Figure 3 and the summary of the results on all four of these parameters of the four PPI networks is presented in Table 1. Despite the degree distribution of these PPI networks being closest to the degree distributions of the corresponding scale-free random networks (see Table 1), the remaining three parameters of PPI networks differ from the scale-free model with two thirds of them being closest to the corresponding geometric random networks. Also, many of these properties of the two fruitfly PPI networks were close to ER, ER-DD, and SF models possibly indicating the presence of noise in these PPI networks (see Supplementary Information). Nevertheless, the high-confidence fruitfly PPI network exhibits some geometric network properties. We expect that increasing confidence in the fruitfly PPI data set will make the structural properties of its PPI network closer to those of the geometric random graphs.

In summary, we have shown compelling evidence that the structure of yeast PPI networks is closer to the geometric random graph model than to the currently accepted scale-free model. Three out of four of the commonly studied statistical properties of global network structure, as well as the newly introduced graphlet frequency parameter describing local structural properties of large networks, of yeast PPI networks were closer to geometric random graphs than to scale-free or Erdös-Rényi random graphs. In addition, despite the noise present in their PPI detection techniques and the lack of independent verification of its PPIs by various labs, fruitfly

PPI networks do show properties of geometric random graphs. It should be noted that cellular processes happen in 3-dimensional space and over time, so it is not surprising that 3- and 4-dimensional geometric graph models seem to provide good fits to these data. In addition, other designed and optimized communication networks, such as wireless multihop networks[30], electrical power-grid and protein structure networks[19], have been modeled by geometric random graphs. Thus, it is plausible that PPI networks, which possibly emerged, similar to the World Wide Web, through stochastic growth processes, but unlike the World Wide Web have gone through millions of years of evolutionary optimization, are better modeled by the geometric random graph model then by the scale-free model (scale-free model seems to be appropriate for networks which have emerged through stochastic growth processes and have not been optimized, such as the World Wide Web). Also, similar to the limited coverage that wireless networks have, currently available PPI data cover only a portion of the interactome. Once a more complete interactome data becomes available, we will be able to validate the correctness of the current and possibly design better models for PPI networks.

*Note: Supplementary information is available and submitted together with this manuscript.*

ACKNOWLEDGMENTS

We thank Rudi Mathon, Gil Prive, Wayne Hayes, and Isidore Rigoutsos for helpful comments and discussions, and Andrew King for implementing some of the random graph generators. Financial support from the Natural Sciences and Engineering Research Council of Canada, the Ontario Graduate Scholarship Program and IBM Canada was gratefully received.

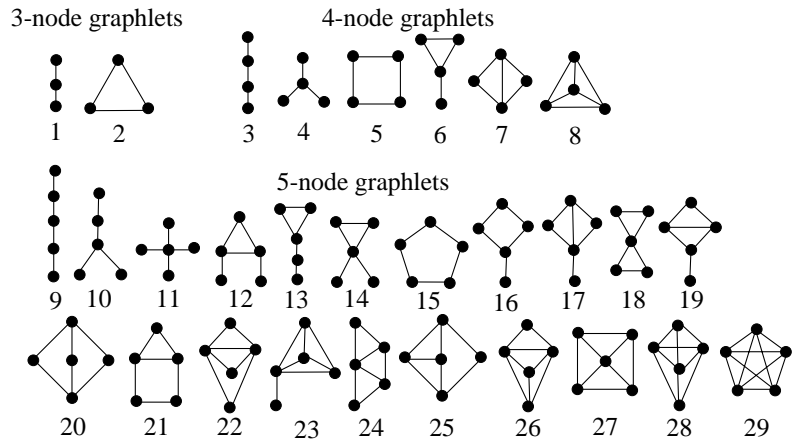

Figure 1: All 3-node, 4-node, and 5-node networks (graphlets), ordered within groups from the least to the most dense with respect to the number of edges when compared to the maximum possible number of edges in the graphlet; they are numbered from 1 to 29.

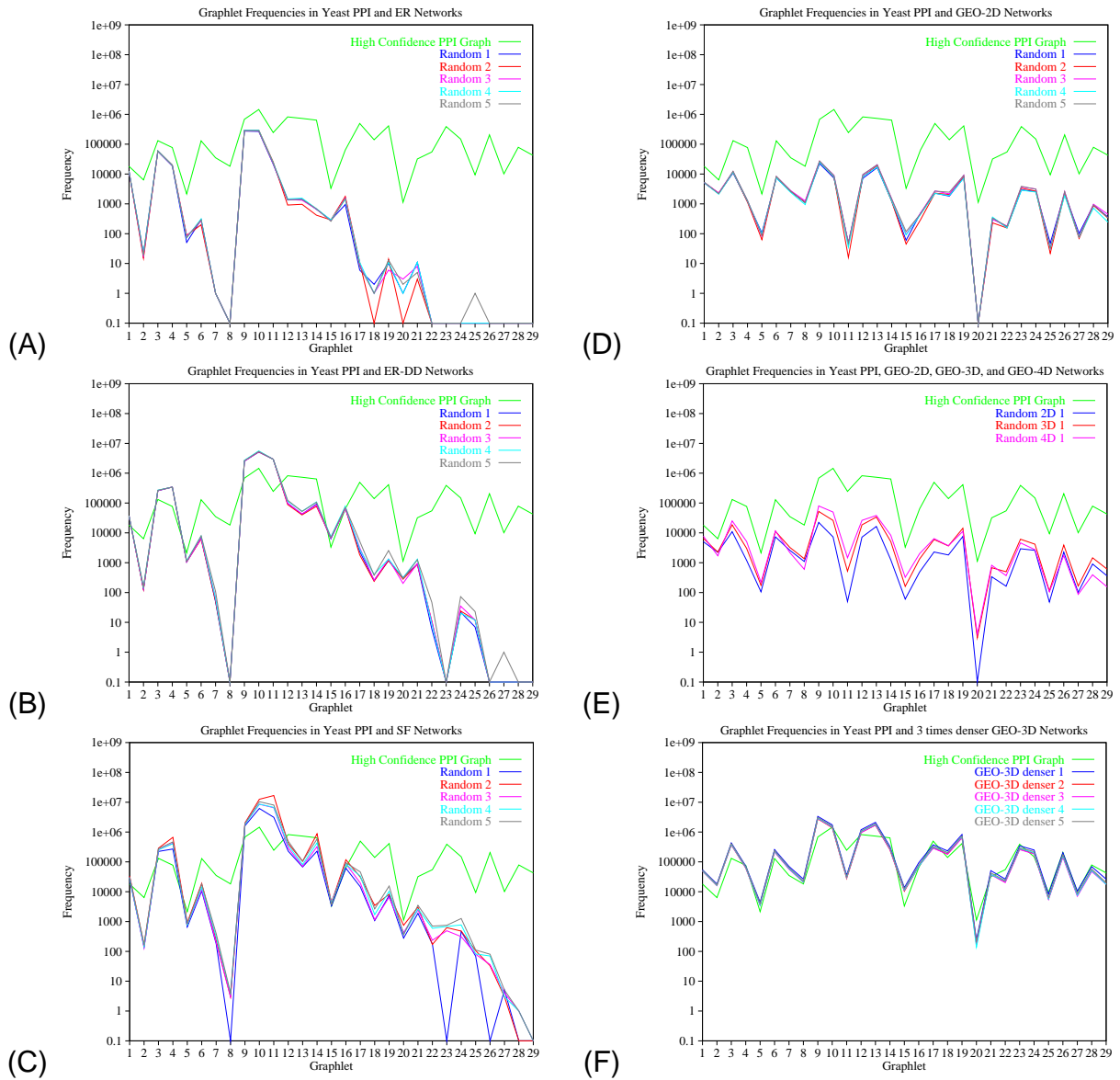

Figure 2: Comparison of graphlet frequencies in the high-confidence *S. cerevisiae* PPI network[29] (green line) with corresponding ER, ER-DD, SF, and GEO random networks. Zero frequencies were replaced by 0.1 for plotting on log-scale. **A.** PPI network *versus* five corresponding ER random networks. **B.** PPI network *versus* five corresponding ER-DD random networks. **C.** PPI network *versus* five corresponding SF random networks. **D.** PPI network *versus* five corresponding GEO-2D random networks. **E.** PPI network *versus* a corresponding GEO-2D, GEO-3D, and GEO-4D random network. **F.** PPI network *versus* five GEO-3D random networks with the same number of nodes and approximately three times as many edges as the PPI network.

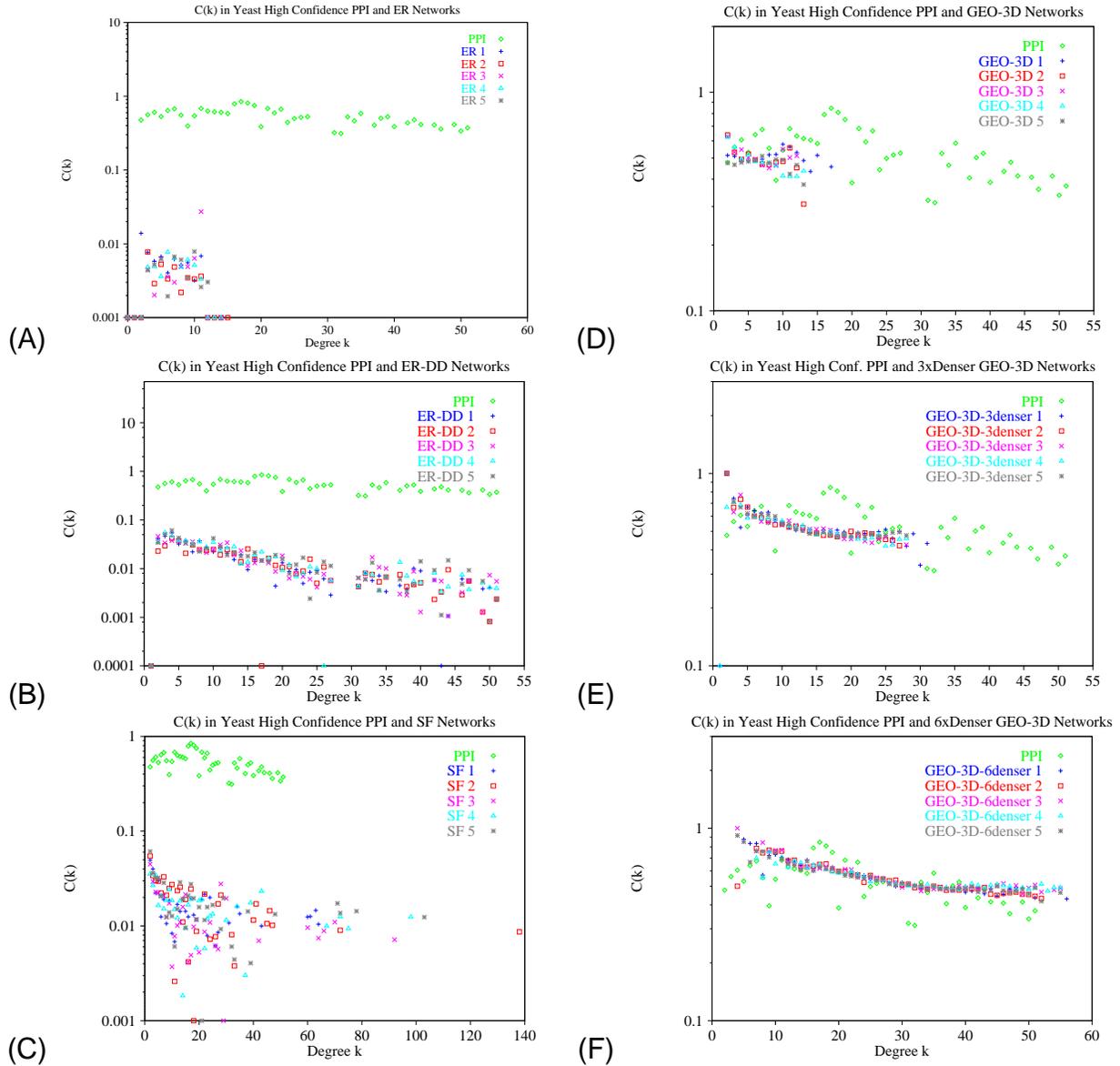

Figure 3: Comparison of average clustering coefficients $C(k)$ of degree $k$ nodes in the high-confidence *S. cerevisiae* PPI network[29] (green dots) with corresponding ER, ER-DD, SF, and GEO random networks. Zero values were replaced by values close to zero to be plotted on the abscissa on log-scale. **A.** PPI network *versus* five corresponding ER random networks. **B.** PPI network *versus* five corresponding ER-DD random networks. **C.** PPI network *versus* five corresponding SF random networks. **D.** PPI network *versus* five corresponding GEO-3D random networks. **E.** PPI network *versus* five corresponding GEO-3D random networks with the same number of nodes and approximately three times as many edges as the PPI network. **F.** PPI network *versus* five corresponding GEO-3D random networks with the same number of nodes and approximately six times as many edges as the PPI network.

|  | Deg. Distrib. | Diameter | Clust. Coeff. $C$ | $C(k)$ |
|---|---|---|---|---|
| Yeast High-Conf. | SF | GEO-3D 3×denser | GEO-4D | GEO-3D 6×denser |
| Yeast 11K | SF | GEO-3D 6×denser | GEO-4D | GEO-3D 6×denser |
| Fruitfly High-Conf. | SF | SF / GEO-3D 6×denser | No Fit | between GEO & ER |
| Fruitfly Larger | SF | ER-DD / SF | SF / ER-DD | SF |

Table 1: Behavior of the degree distribution, diameter, and clustering coefficients $C$ and $C(k)$ of four PPI networks (yeast high-confidence and top 11000 interaction[29], and fruitfly high-confidence and larger[27]), when compared to the corresponding ER, ER-DD, SF, and GEO random networks. The values in the boxes represent the random graph model which provides the best fit for the PPI networks with respect to the given parameter. For example, "SF" means that the degree distribution for a PPI networks is closer to the degree distributions of corresponding scale-free random networks then to the corresponding ER, ER-DD, and GEO random networks. "GEO-3D 3×denser" and "GEO-3D 6×denser" denote GEO-3D networks with the same number of nodes, but about three and six time as many edges as the corresponding PPI networks, respectively.

# A Supplementary Methods and Discussion

## A.1 Models of Large Networks

The earliest model of large networks was introduced by Erdös and Rényi in late 50s and early 60s [1, 2, 3]. They introduced the *random graph* model and initiated a large area of research, a good survey of which can be found in Bollobas's book [31]. There are several versions of this model out of which the most commonly studied is the one based on the principle that in a graph on $n$ nodes, every edges is present with probability $p$ and absent with probability $1 - p$; this graph model is commonly denoted by $G_{n,p}$. Another commonly used random graph model, denoted by $G_{n,m}$, is based on the principle that a graph on $n$ nodes and $m$ edges is chosen uniformly at random amongst all graphs on $n$ nodes and $m$ edges, or equivalently, a set of $m$ edges of the graph is chosen uniformly at random among all possible $\binom{\binom{n}{2}}{m}$ sets of edges. These two models behave similarly when $p \approx \frac{m}{\binom{n}{2}}$. We used the $G_{n,m}$ model to construct ER graphs corresponding to the PPI networks (details are below). Many of the properties of random graphs can be calculated exactly in the limit of large $n$ [31].

Since properties of the random graph model deviate from those of most real-world networks, several new network models have recently been introduced. To capture the scale-free character of real-world networks, the random graph model was modified to allow for arbitrary degree distributions while keeping all other aspects of the random graph model. This model is called *generalized random graph* model and finding properties of these graphs has been an active area of research [32, 33, 34, 35, 36, 37, 38, 39]. The ER-DD random networks which we constructed to have the same number of nodes, edges, and the degree distributions as the PPI networks belong to this network model. We surveyed some of the main results on these graphs in [40].

The *scale-free* network model, characterized by a small number of very highly connected nodes, has received a lot of attention in recent years. Barabási, Albert, and Jeong [5, 41] showed that a heavy-tailed degree distribution emerges automatically from a stochastic growth model in which new nodes are added continuously and they preferentially attach to existing nodes with probability proportional to the degree of the target node. That is, high-degree nodes become of even higher degree with time and the resulting degree distribution is $P(k) \approx k^{-\gamma}$. We constructed the SF networks corresponding to PPI networks using this network model (more details are given below). Many real-world networks have power-law degree distributions, such as for example, the Internet backbone [42], metabolic reaction networks [43], the telephone call graph [44], and the World Wide Web [45]; degrees of all of these networks decay as a power law $P(k) \approx k^{-\gamma}$, with the exponent $\gamma \approx 2.1 - 2.4$. Thus, a large body of research on theoretical and experimental results on this network model has appeared recently [46, 37, 38, 47, 9, 48, 49, 50, 51, 45, 51, 52, 53, 54, 43, 55].

A *geometric graph* $G(V, r)$ with *radius* $r$ is a graph with the node set $V$ of points in a metric space and the edge set $E = \{\{u, v\} | (u, v \in V) \land (0 < \|u - v\| \leq r)\}$, where $\| \cdot \|$ is an arbitrary distance norm in this space. That is, points in a metric space correspond to nodes, and two nodes are adjacent if the distance between them is at most $r$. Often, two dimensional space is considered, containing points in the unit square $[0, 1]^2$ or unit disc, and $0 < r < 1$ [56, 57], with the distance norms being $l_1$ (Manhattan distance), $l_2$ (Euclidean distance), or $l_\infty$ (Chessboard distance). The distance between two points $(x_1, y_1)$ and $(x_2, y_2)$ is $|x_1 - x_2| + |y_1 - y_2|$ in $l_1$ norm, $\sqrt{(x_1 - x_2)^2 + (y_1 - y_2)^2}$ in $l_2$ norm, and $max(|x_1 - x_2|, |y_1 - y_2|)$ in $l_\infty$ norm. A *random geometric graph* $G(n, r)$ is a geometric graph with $n$ nodes which correspond to $n$ independently and uniformly randomly distributed points in a metric space. Many properties of these graphs have been explored when $n \to \infty$ [28]. Similar to Erdös-Rényi random graphs, certain properties of these graphs also appear suddenly when a specific threshold is reached. We used 2-, 3-, and 4-dimensional "squares" and the Euclidean distance measure to construct GEO-2D, GEO-3D, and GEO-4D graphs corresponding to PPI networks (more details are given below).

## A.2 Graphlet Analysis

We analyzed graphlet frequencies of four PPI networks: (1) the high-confidence yeast *S. cerevisiae* PPI network involving 2455 interactions amongst 988 proteins [29]; (2) the yeast *S. cerevisiae* PPI network involving 11000 interactions amongst 2401 proteins [29] (these are the top 11000 interactions in von Mering *et al.* classification [29]); (3) the high-confidence fruitfly *D. melanogaster* PPI network involving 4637 interactions amongst 4602 proteins [27]; (4) and the entire fruitfly *D. melanogaster* PPI network as published in [27] involving 20007 interactions amongst 6985 proteins which includes low confidence interactions. All graphlet counts were obtained by the exhaustive graphlet search algorithm going through node adjacency lists rather than the adjacency matrix for increased efficiency. LEDA library for combinatorial and geometric computing was used in its implementation [58]. In graphlet frequency figures in the paper and in the supplementary graphlet frequency figures, graphlet numbers on the abscissae are ordered as in Figure 1 in the paper, so that the node and edge numbers of the graphlets they represent are monotonically increasing; all zero graphlet frequencies are approximated by $0.1$ for plotting them on log-scale. The same replacement of 0 graphlet counts by 0.1 was done for calculating the distances between networks (for taking logarithm in the distance formula described in the paper).

To evaluate how well the PPI networks fit different network models with respect to their graphlet frequencies, we compared frequencies of the 29 3-5-node graphlets in PPI networks against their frequencies in different types of random networks. For each of the PPI networks we constructed five different random graphs belonging to each of the following four random graph models: (1) Erdös-Rényi $G_{n,m}$ random graphs having the same number of nodes and edges as the corresponding PPI networks (denoted by ER); (2) Erdös-Rényi $G_{n,m}$ random graphs having the same number of nodes, edges, and the degree distribution as the corresponding PPI networks (denoted by ER-DD); (3) scale-free random graphs having the same number of nodes and the number of edges within 1% of those of the corresponding PPI networks (denoted by SF); (4) 2-, 3-, and 4-dimensional geometric random graphs with the number of nodes and the number of edges within 1% of those of the corresponding PPI networks (denoted by GEO-2D, GEO-3D, and GEO-4D, respectively). Note that it is enough to compare graphlet frequencies of the PPI networks against a very small number of random graphs of each of these types because graphs belonging to each network model have almost identical graphlet frequency distribution (this can be experimentally observed in Supplementary Figures 5 - 6 and Supplementary Table 3 and also theoretically proved).

### A.2.1 Construction of Model Networks

We tested graphlet frequencies of five different $G_{n,m}$ graphs for $n = 988$ nodes and $m = 2455$ edges with those of five different $G_{n,p}$ graphs with the same $n$ and $p = \frac{m}{\binom{n}{2}}$ and obtained identical frequency distributions. Thus, we used the $G_{n,m}$ model in our analysis. We used the random graph generation function from LEDA library to construct ER graphs of $G_{n,m}$ type corresponding to PPI networks. The ER-DD random graphs were constructed by generating $n$ nodes, where $n$ is equal to the number of nodes in a PPI network, assigning the degree sequence of a PPI network to these generated nodes, sorting the nodes by degree in decreasing order, and generating edges from the sorted nodes towards randomly selected nodes while preserving the assigned degree distribution. Since this process yields some impossible edge assignments, we repeated this process several times until it yielded the desired graph; it took between 7 and 15 experiments to generate each of the desired ER-DD graphs.

Scale-free random graphs $G(V, E)$ were generated in the following way. Let $k = \frac{|E|}{|V|}$, and let an integer $i$ be such that $i \leq k \leq i + 1$. We start the graph construction with an independent set of size $i$. We add a node to this graph and connect it with the $i$ nodes of the initial independent set. We add subsequent nodes and connect them with either $i$ or $i + 1$ other nodes in the graph with probabilities $i + 1 - k$ and $k - i$ respectively; attachment is preferential (a new node is more likely to be attached to a high-degree than to a low-degree node) and directly proportional to the degree of the node that a new node is being attached to.

We constructed 2-, 3-, and 4-dimensional geometric random graphs with $l_2$ (Euclidean) norm corresponding to each of the above mentioned PPI networks. The following parameters were used for their construction: a $95 \times 95$ square and radius 4 for the GEO-2D graphs corresponding to the high confidence yeast PPI networks; a $150 \times 150$ square and radius 5.3 for the GEO-2D graphs corresponding to the "top 11000" yeast PPI network; a $250 \times 250$ square and radius 3 for GEO-2D random graphs corresponding to the high confidence fruitfly PPI network; a $310 \times 310$ square and radius 5 for GEO-2D graphs corresponding to the entire fruitfly PPI network as published in [27]; a $50 \times 50 \times 50$ cube and radius 5.55 for GEO-3D graphs corresponding to the high confidence yeast PPI network; a $50 \times 50 \times 50$ cube and radius 5 for GEO-3D graphs corresponding to the "top 11000" yeast PPI network; a $84 \times 84 \times 84$ cube and radius 4 for GEO-3D graphs corresponding to the high confidence fruitfly PPI network; a $67 \times 67 \times 67$ cube and radius 4 for GEO-3D graphs corresponding to the "entire" fruitfly PPI network; a $30 \times 30 \times 30 \times 30$ cube and radius 5.74 for GEO-4D graphs corresponding to the high confidence yeast PPI network; a $30 \times 30 \times 30 \times 30$ cube and radius 5.25 for GEO-4D graphs corresponding to the "top 11000" yeast PPI network; a $30 \times 30 \times 30 \times 30$ cube and radius 3 for GEO-4D graphs corresponding to the high confidence fruitfly PPI network; a $35 \times 35 \times 35 \times 35$ cube and radius 4.13 for GEO-4D graphs corresponding to the "entire" fruitfly PPI network. GEO-3D graphs with the same number of nodes, but approximately three times as many edges as the high-confidence yeast PPI network were obtained from a $50 \times 50 \times 50$ cube and radius 8 and they contained 7126, 6789, 6805, 6824, and 7004 edges respectively. GEO-3D graphs with the same number of nodes, but approximately six times as many edges as the high-confidence yeast PPI network were obtained from a $50 \times 50 \times 50$ cube and radius 10.3 and they contained

14257, 14506, 14177, 14185 and 14302 edges respectively.

**A.2.2 Graphlet Frequency Results**

We first established that graphlet frequencies of the two yeast PPI networks are highly correlated (see Supplementary Figure 4 A and B). To quantify the correlation between the graphlet frequencies for these two and other networks, we used the following simple method (also described in the paper). We first normalized graphlet frequencies of each graph in the following way: let $N_i(G)$ be the number of graphlets of type $i$ ($i \in \{1, \ldots, 29\}$) of graph $G$; let the total number of graphlets of $G$ be denoted by $T(G) = \sum_{i=1}^{29} N_i(G)$; we computed normalized values of frequencies, $F_i(G) = -log(\frac{N_i(G)}{T(G)})$. Note that we use $log$ because we are interested in fractional (or percentage) differences between graphlet frequencies, and we change its sign to make its value positive and plotting them more intuitive. Then we measured differences between normalized graphlet frequencies of graphs $G$ and $H$ using the *distance* function $D(G, H) = \sum_{i=1}^{29} |F_i(G) - F_i(H)|$; the smaller the value of the distance function $D(G, H)$ is, the more correlated the graphlet frequencies of graphs $G$ and $H$ are. The plots of normalized graphlet frequencies for the two yeast PPI networks are presented in Supplementary Figure 4 B; the distance between these two PPI networks is 21.336207. Note that the distance between the two fruitfly PPI networks is about three times as high, 66.298842, indicating presence of noise in the larger network, as expected; this can also be seen in Supplementary Figure 4 C and D, which shows non-normalized and normalized graphlet frequencies in the two fruitfly PPI networks, respectively.

Non-normalized plots of graphlet frequencies for yeast PPI networks and their corresponding ER, ER-DD, and SF random networks are presented in Supplementary Figure 5. Clearly, these plots show that graphlet frequencies of the two yeast PPI networks are far from the graphlet frequencies of the corresponding ER, ER-DD, and SF random networks. This can also be seen from Supplementary Table 3, which shows distances between PPI networks and their corresponding random networks computed by the formula given above. In contrast, the graphlet frequencies of yeast PPI networks and their corresponding 2-dimensional geometric random graphs are very close, as presented in Supplementary Figure 6 A and B, and in Supplementary Table 3. For example, distances between graphlet frequencies of the high confidence yeast PPI network and five corresponding SF random networks are between 125.468572 and 142.916451, while the distances between this PPI network and five corresponding GEO-2D graphs are about 3.5 times smaller, i.e., between 35.466770 and 38.960880. The correlation is even stronger between yeast PPI networks and their corresponding 3- and 4- dimensional geometric random graphs, as presented in Supplementary Figure 6 C and D and Supplementary Table 3. Furthermore, the correlation between the yeast high confidence PPI network and 3-dimensional geometric random graphs with the same number of nodes, but about three times as many edges as the high confidence PPI network is particularly striking (Supplementary Figure 6 E and Supplementary Table 3) and so is the correlation between this PPI network and GEO-3D random graphs with the same number of nodes but about 6 times as many edges as the PPI network (Supplementary Figure 6 F and Supplementary Table 3); the distance between the PPI and these random graphs is even lower than the distance between the two yeast PPI networks that we analyzed. Note that we constructed the GEO-3D random graphs which are 6 times denser than the yeast high-confidence PPI network (in terms of the number of edges) in order to obtain similar maximum degrees of the PPI and these model networks (more explanation is given in the paper). We expect that once the complete PPI network for yeast becomes available, it will be much denser than the one we are working with today and it will likely have properties of 3- or 4-dimensional geometric random graphs. This hypothesis seems plausible since it is based on examining local structural properties of PPI networks (more details are given in the paper), and also because processes in the cell, including protein-protein interactions, are happening in the 3-dimensional space; cellular processes happen over time, often in a specific time sequence, so adding time as the fourth dimension sounds like a reasonable model.

The results of the graphlet frequency analysis of the fruitfly PPI networks are presented in Supplementary Figures 7 and 8, and in Supplementary Table 3. Even though the fruitfly high confidence PPI network fits the 4-dimensional geometric random graph model about 1.3 times better then the next closest model, which is the ER model, and about 3.2 times better than the SF model with respect to graphlet frequencies (Supplementary Table 3), the difference in fits is not as striking as for the two yeast PPI networks. One explanation for this may be that, since this is the first publicly available fruitfly PPI data set obtained from cDNAs representing each predicted transcript of the fruitfly genome [27], what we are observing is a random sample of the full fruitfly PPI network; by "full" PPI network we mean the PPI network containing all proteins and all PPIs from all cell types in an organism. An alternative explanation may be that, since *D. melanogaster* is a multicellular organism while yeast *S. cerevisiae* consists of a single cell, a different model may be required for the complete fruitfly PPI network, and yet different models may be needed for PPI networks of different fruitfly cells that belong to different tissues. If this is the case, it is reasonable to expect to observe a common PPI network model for all multicellular organisms. Also, the larger fruitfly PPI network, with about 77% of its edges corresponding to lower confidence interactions [27], is closest to the scale-free model (Supplementary Figure 7 F and Supplementary Table 3). This is one of the reasons why we believe that the scale-free properties that have been observed in PPI networks are due to the presence of large amount of noise in these networks; we believe that the true structure of PPI networks is closer to the geometric graph model than to the scale-free model.

The graphlet frequency parameter is robust to random perturbations (Supplementary Table 2 and Supplementary Figures 9, 10, and 11). We perturbed the high-confidence yeast PPI network by randomly adding, deleting, and rewiring 10, 20, and 30 percent of its edges and computed distances between the perturbed networks and the PPI network by using the distance function defined above (and in the paper). We constructed five perturbed networks in each of these nine categories (45 perturbed networks in total). We found the exceptional robustness of the graphlet frequency parameter to random additions of edges very encouraging, especially in light of the currently available PPI networks containing many false negatives (missing edges). In particular, additions of 30% of edges resulted in networks which were about 21 times closer to the PPI network than the corresponding SF random networks (the distances between the PPI and the perturbed PPI networks with additions of 30% of edges were between 5.788158 and 7.149713, while the distance between the PPI and the corresponding SF random networks were between 125.468572 and 142.916451, as shown in Supplementary Tables 2 and 3). We also found that graphlet frequencies were fairly robust to random edge deletions and rewirings (deletions and rewirings of 30% of edges resulted in networks which were about 5.8 and 5.4 times closer to the PPI network than the corresponding SF random networks, respectively), which further increases our confidence in PPI networks having geometric properties despite the presence of false positives in the data.

## A.3 Standard Global Network Parameters

### A.3.1 Definitions

The most commonly studied statistical properties of large networks measuring their global structure are the degree distribution, network diameter, and clustering coefficients, defined as follows (also defined in the paper). The *degree* of a node is the number of edges (connections) incident to the node. The *degree distribution*, $P(k)$, describes the probability that a node has degree $k$. This network property has been used to distinguish between different network models; in particular, Erdös-Rényi random networks have a Poisson degree distribution, while *scale-free* networks have a power-law degree distribution $P(k) \sim k^{-\gamma}$, where $\gamma$ is a positive number. The smallest number of links that have to be traversed to get from node $x$ to node $y$ in a network is called the *distance* between nodes $x$ and $y$ and a path through the network that achieves this distance is called a *shortest path* between $x$ and $y$. The average of shortest path lengths over all pairs of nodes in a network is called the network *diameter*. (Note that in classical graph theory, the diameter is the maximum of shortest path lengths over all pairs of nodes in the network [12]; we do not use this definition.) This network property also distinguishes different network models: for example, the diameter of Erdös-Rényi random networks on $n$ nodes is proportional to $\log n$, the network property often referred to as the *small-world* property; the diameters of scale-free random networks with degree exponent $2 < \gamma < 3$, which have been observed for most real-world networks, are *ultra-small* [13, 14], i.e., proportional to $\log \log n$. The *clustering coefficient of node* $v$ in a network is defined as $C_v = \frac{2e_1}{n_1(n_1-1)}$, where $v$ is linked to $n_1$ neighboring nodes and $e_1$ is the number of edges amongst the $n_1$ neighbors of $v$. The average of $C_v$ over all nodes $v$ of a network is the *clustering coefficient* $C$ of the whole network and it measures the tendency of the network to form highly interconnected regions called clusters. The average clustering coefficient of all nodes of degree $k$ in a network, $C(k)$, has been shown to follow $C(k) \sim k^{-1}$ for many real-world networks indicating a network's hierarchical structure [15, 6]. Many real world networks have been shown to have high clustering coefficients and to exhibit small-world and scale-free properties.

### A.3.2 Results on PPI and Model Networks

We obtained the degree distributions, diameters, and clustering coefficients for the above four PPI networks, as well as for the recently published *C. elegans* PPI network [59] involving 5363 interactions between 3115 proteins. We did not obtain graphlet frequencies for this data set because it contains a large number of highly connected nodes which are in close proximity of each other. This makes it infeasible to find graphlets in this PPI network using the standard, brute force exhaustive search technique, which we have used to find graphlets in the yeast and fruitfly PPI networks, as well as in all of their corresponding model networks.

We confirmed that degree distributions of all of these PPI networks approximately follow power law. In Supplementary Figure 12 we present power law functions fitted to the degree distributions of these five PPI networks. The degree distributions of the fruitfly PPI networks (Supplementary Figure 12 C and D) deviate the most from power-law functions. Note that the exponents $\gamma$ of the fitted power law functions $P(x) = x^{-\gamma}$ are between $1.2$ and $1.7$ for most of these PPI networks (the only exception is the high confidence fruitfly PPI network, but the approximation of its degree distribution by $x^{-2.4}$ is very poor). This deviates from what was observed in many other real-world networks, including metabolic pathway networks of 43 different organisms [43, 23], where these exponents were between 2 and 3. An illustration of the degree distributions of the PPI networks against the degree distributions of the corresponding model networks is presented in Supplementary Figure 13.

When calculating a network diameter, we used the standard method of considering only the lengths of shortest paths between reachable pairs of nodes, i.e., pairs of nodes which are in the same connected component of the network. Diameters of the PPI networks and the corresponding random networks are presented in Supplementary Table 4. We observed that the diameters of yeast PPI networks are closest to the diameters of GEO-3D networks which are 3 or 6 times denser than the corresponding PPI networks, while diameters of fruitfly and worm PPI networks are closest to the diameters of the corresponding SF networks (the only exception is the diameter of the larger fruitfly PPI network which is slightly closer to the diameters of ER-DD than to the diameters of SF networks; Supplementary Table 4). However, once a network is dense enough to have most of its nodes in the same connected component, with increasing density of GEO-3D (or any other type of) networks on the same number of nodes, their diameters decrease (this is theoretically expected and can also be experimentally observed in Table Supplementary 4), so it would not be hard to construct GEO-3D networks with the same number of nodes but more edges than the fruitfly and the worm PPI networks have to achieve a closer fit of the diameters of the PPI and the GEO-3D model networks. Thus, GEO graphs *do* model PPI networks with respect to this network parameter.

Clustering coefficients of the PPI and the corresponding model networks are presented in Supplementary Table 5. Clustering coefficients of the two yeast PPI networks are orders of magnitude larger than the clustering coefficients of the corresponding ER, ER-DD, and SF random networks while they are in agreement with the clustering coefficients of the corresponding GEO graphs (Supplementary Table 5). This can not be said for the other three PPI networks. In particular, the clustering coefficient of the high-confidence fruitfly PPI network differs by an order of magnitude from all model networks, while the clustering coefficients of the larger fruitfly and the worm PPI networks are closest to the clustering coefficients of the corresponding SF random networks; we believe that this may be due to the large amount of noise being present in these PPI networks.

We measured the average clustering coefficients of all nodes of degree $k$ in a network, $C(k)$, for the above mentioned PPI and their corresponding model networks (Supplementary Figures 14 – 18); zero values are replaced with values close to zero and plotted along the abscissa in these supplementary figures. High correlations between $C(k)$ of the two yeast PPI and their corresponding GEO-3D networks (especially the corresponding GEO-3D networks which are about 6 times denser than the PPI networks, Supplementary Figures 14 F and 15 F) and the lack of such correlation with other model networks is blindingly obvious (Supplementary Figures 14 and 15). However, the values of $C(k)$ are much lower for the other three PPI networks and do not seem to correlate with $C(k)$ of any of the corresponding model networks (Supplementary Figures 16 – 18); we find these values of $C(k)$ of PPI networks to be unrealistically small for real-world networks and believe they are an artifact of the lack of PPI data for fruitfly and worm (false negatives). Also, we observed a lack of scaling of $C(k) \sim k^{-1}$ in all of these PPI networks (Supplementary Figures 14 – 18).

# B  Supplementary Figures

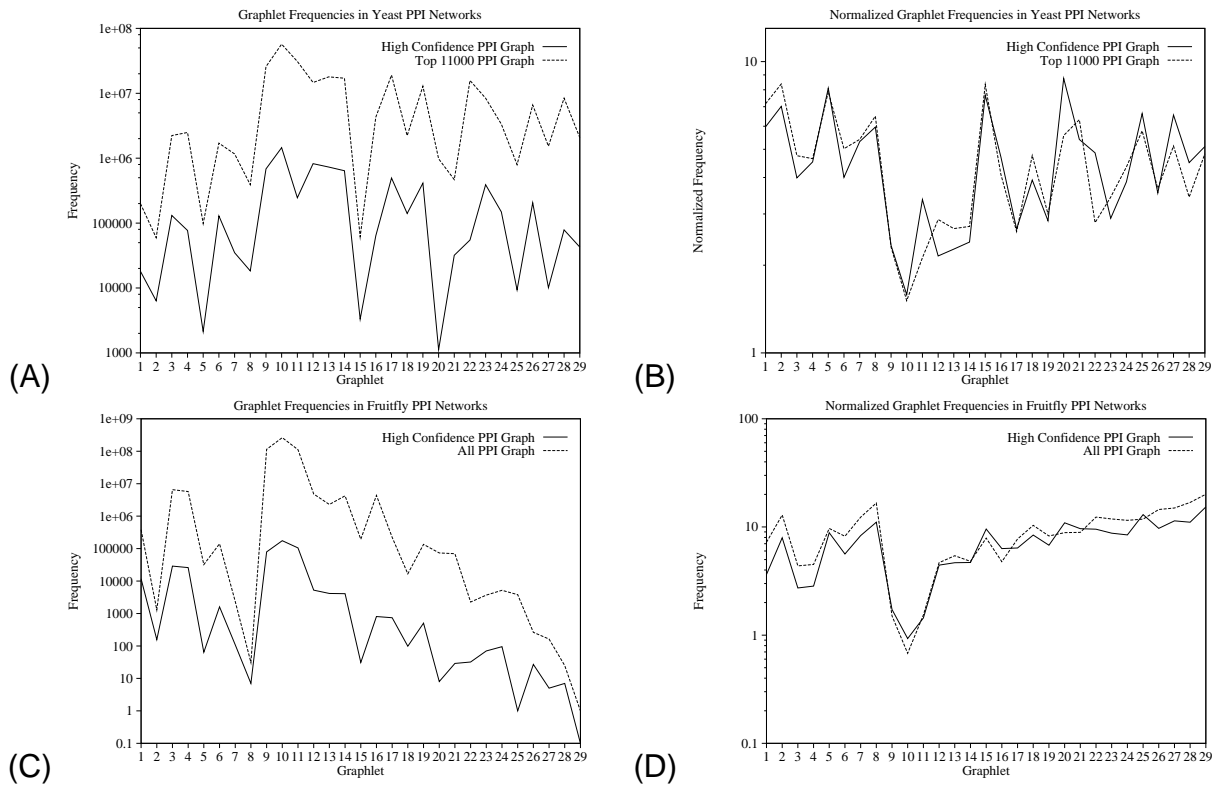

Figure 4: Graphlet frequencies in the two yeast PPI networks [29] and in the two fruitfly PPI networks [27]: **A.** Non-normalized frequencies for the two yeast PPI networks. **B.** Normalized frequencies for the two yeast PPI networks. **C.** Non-normalized frequencies for the two fruitfly PPI networks. **D.** Normalized frequencies for the two fruitfly PPI networks.

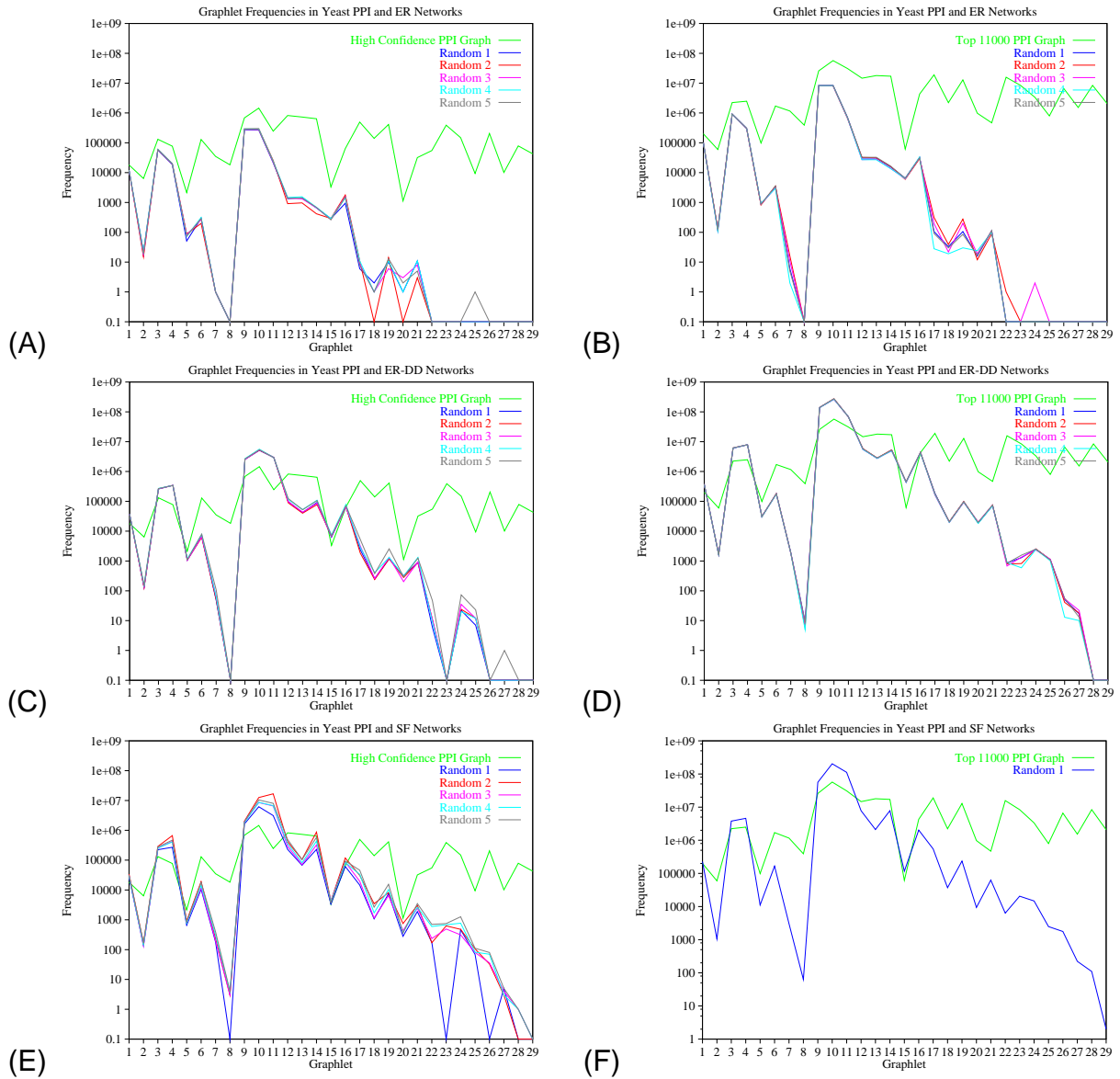

Figure 5: Comparison of frequencies of graphlets in the *S. cerevisiae* PPI networks with ER, ER-DD, and SF random graphs: **A.** High confidence PPI network *versus* the corresponding ER random graphs. **B.** Top 11000 PPI network *versus* the corresponding ER random graphs. **C.** High confidence PPI network *versus* the corresponding ER-DD random graphs. **D.** Top 11000 PPI network *versus* the corresponding ER-DD random graphs. **E.** High confidence PPI network *versus* the corresponding SF random graphs. **F.** Top 11000 PPI network *versus* the corresponding SF random graphs.

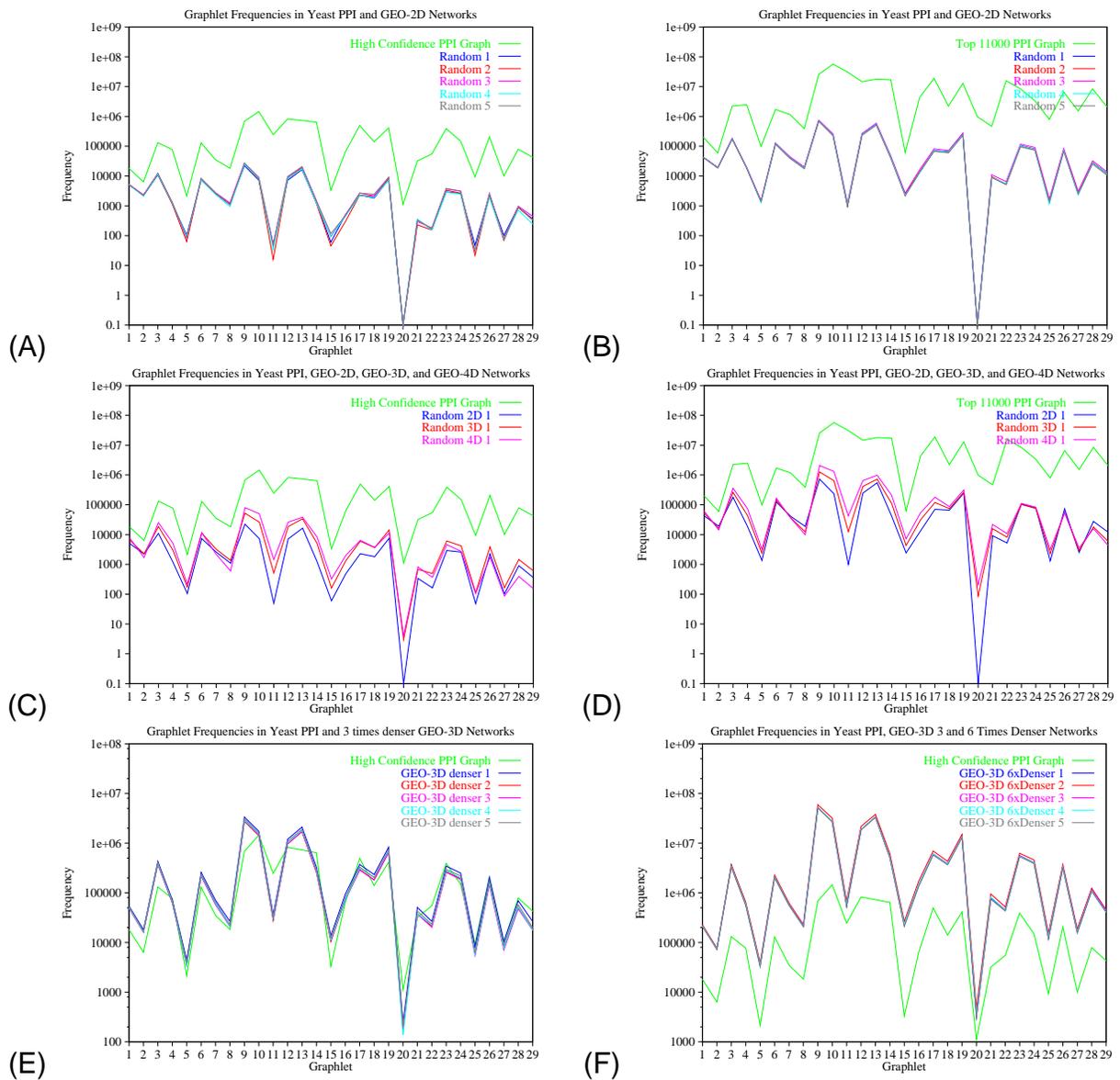

Figure 6: Comparison of frequencies of graphlets in the *S. cerevisiae* PPI networks [29] and geometric random graphs: **A.** High confidence PPI network *versus* the five corresponding GEO-2D graphs. **B.** Top 11000 PPI network *versus* the five corresponding GEO-2D graphs. **C.** High confidence PPI network *versus* a 2-, 3-, and 4-dimensional geometric random graph. **D.** Top 11000 PPI network *versus* a 2-, 3-, and 4-dimensional geometric random graph. **E.** High confidence PPI network *versus* five GEO-3D graphs with the same number of nodes, but approximately three times as many edges as the PPI network. **F.** High-confidence *S. cerevisiae* PPI networks [29] *versus* GEO-3D networks which are about 6 times as dense as the PPI network.

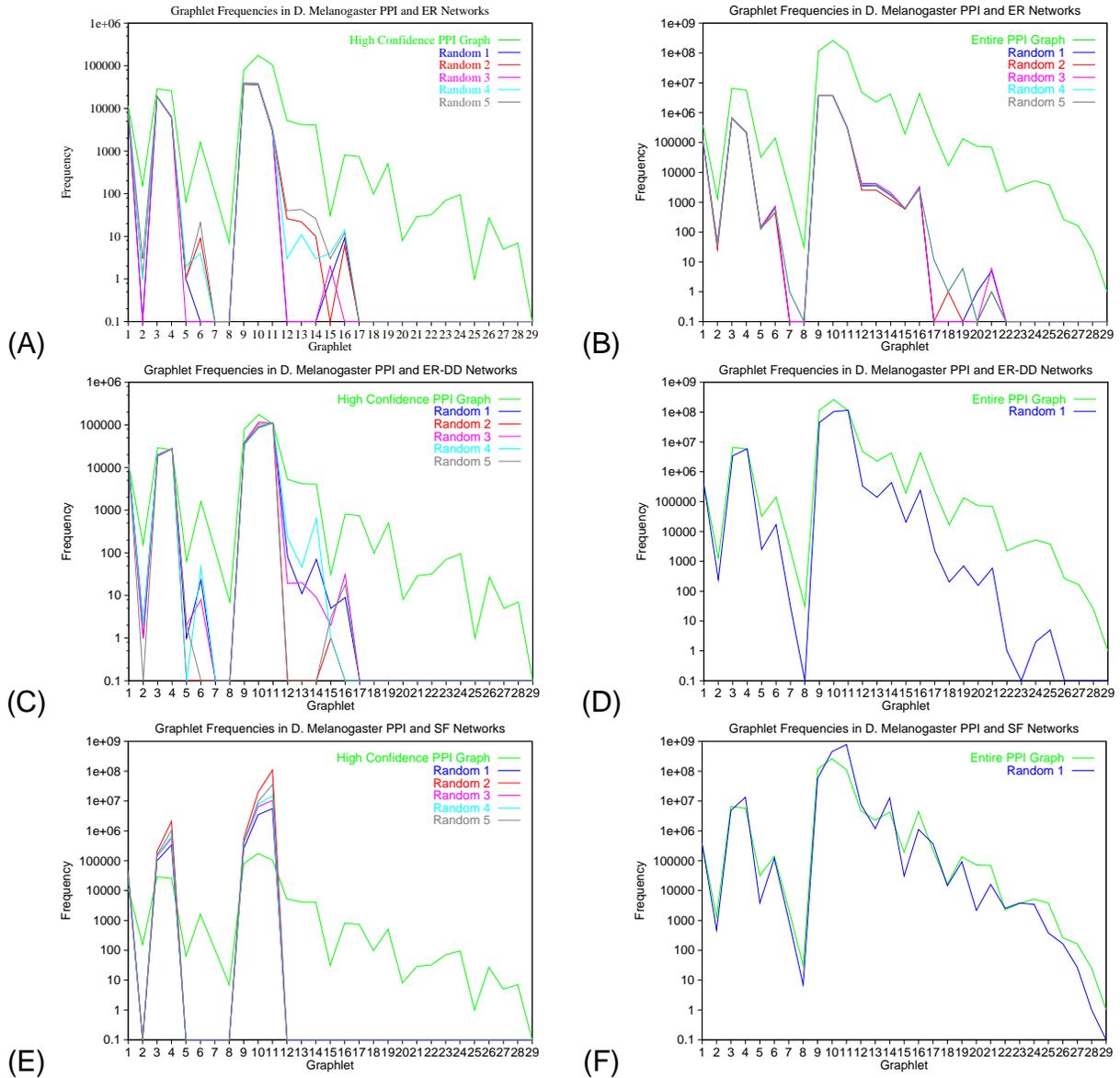

Figure 7: Comparison of frequencies of graphlets in the *D. melanogaster* PPI networks [27] with ER, ER-DD, and SF random graphs: **A.** High confidence fruitfly network *versus* the corresponding ER random graphs. **B.** Entire currently available fruitfly PPI network *versus* the corresponding ER random graphs. **C.** High confidence fruitfly PPI network *versus* the corresponding ER-DD random graphs. **D.** Entire currently available fruitfly PPI network *versus* the corresponding ER-DD random graphs. **E.** High confidence fruitfly PPI network *versus* the corresponding SF random graphs. **F.** Entire currently available fruitfly PPI network *versus* the corresponding SF random graphs.

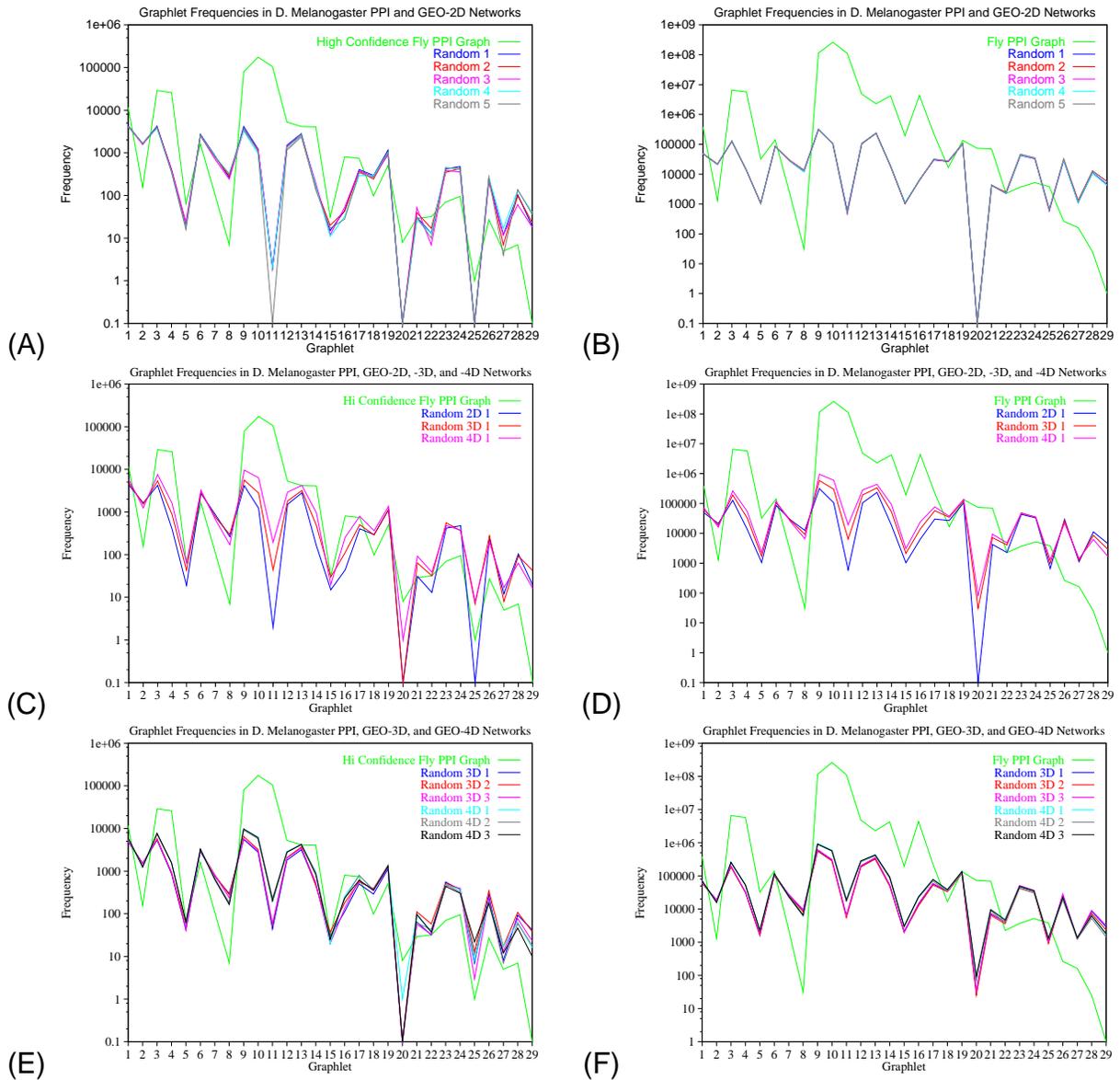

Figure 8: Comparison of frequencies of graphlets in *D. melanogaster* PPI networks [27] with geometric random graphs: **A.** High confidence fruitfly PPI network *versus* GEO-2D random graphs. **B.** Entire currently available fruitfly PPI network *versus* GEO-2D random graphs. **C.** High confidence fruitfly PPI network *versus* GEO-2D, GEO-3D, and GEO-4D random graphs. **D.** Entire currently available fruitfly PPI network *versus* GEO-2D, GEO-3D, and GEO-4D random graphs. **E.** High confidence fruitfly PPI network *versus* three GEO-3D random graphs and three GEO-4D random graphs. **F.** Entire currently available fruitfly PPI network *versus* three GEO-3D random graphs and three GEO-4D random graphs.

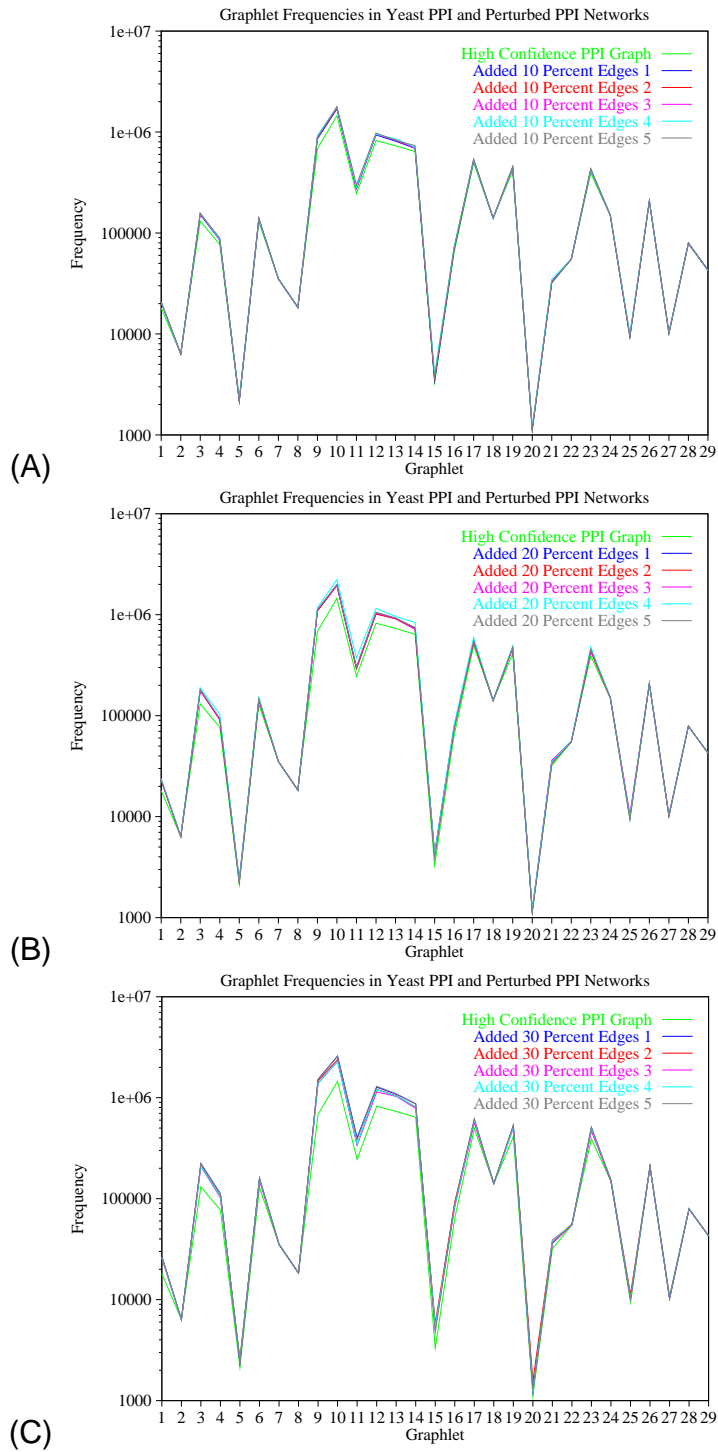

Figure 9: Comparison of graphlet frequencies in the high confidence yeast PPI network with networks obtained by adding edges at random to the PPI network: **A.** five different networks with 10% of edges added. **B.** five different networks with 20% of edges added. **C.** five different networks with 30% of edges added.

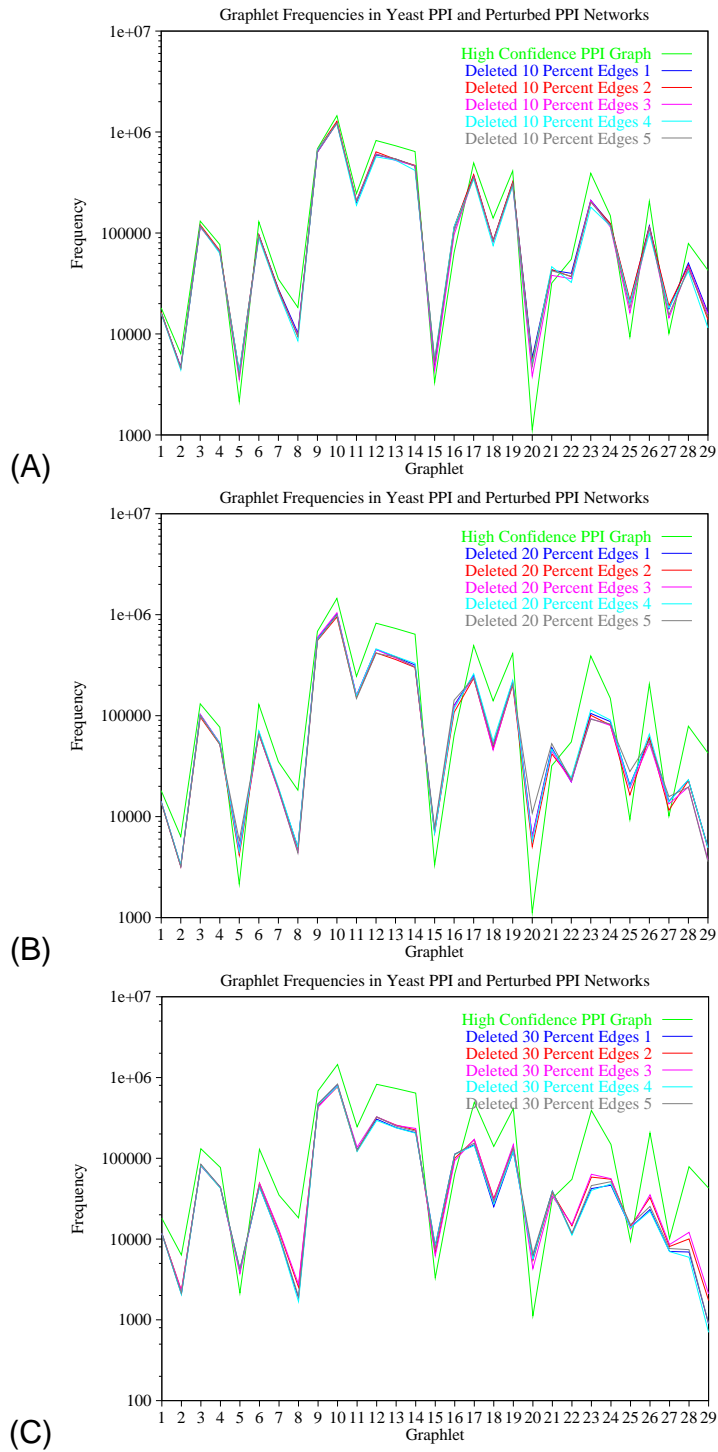

Figure 10: Comparison of graphlet frequencies in the high confidence yeast PPI network with networks obtained by deleting edges at random from the PPI network: **A.** five different networks with $10\%$ of edges deleted. **B.** five different networks with $20\%$ of edges deleted. **C.** five different networks with $30\%$ of edges deleted.

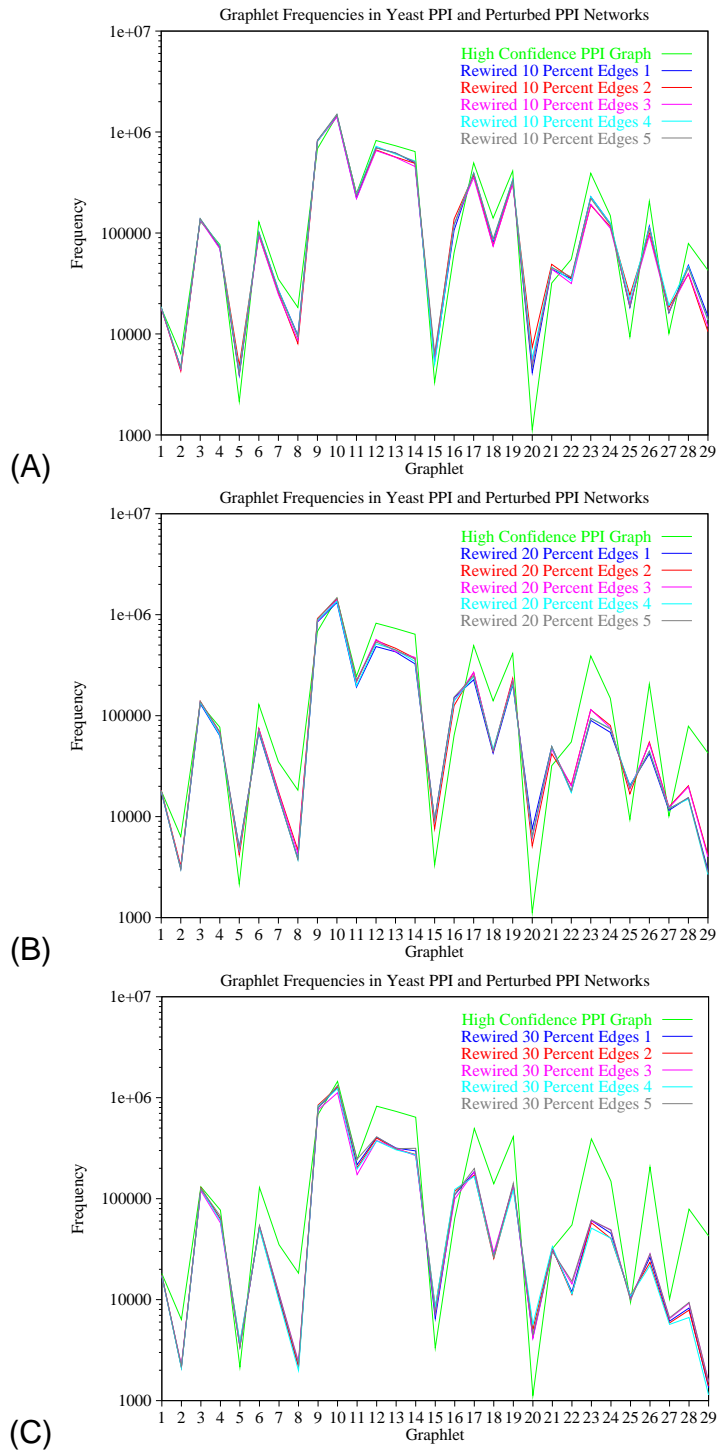

Figure 11: Comparison of graphlet frequencies in the high confidence yeast PPI network with networks obtained by rewiring edges at random in the PPI network: **A.** five different networks with $10\%$ of edges rewired. **B.** five different networks with $20\%$ of edges rewired. **C.** five different networks with $30\%$ of edges rewired.

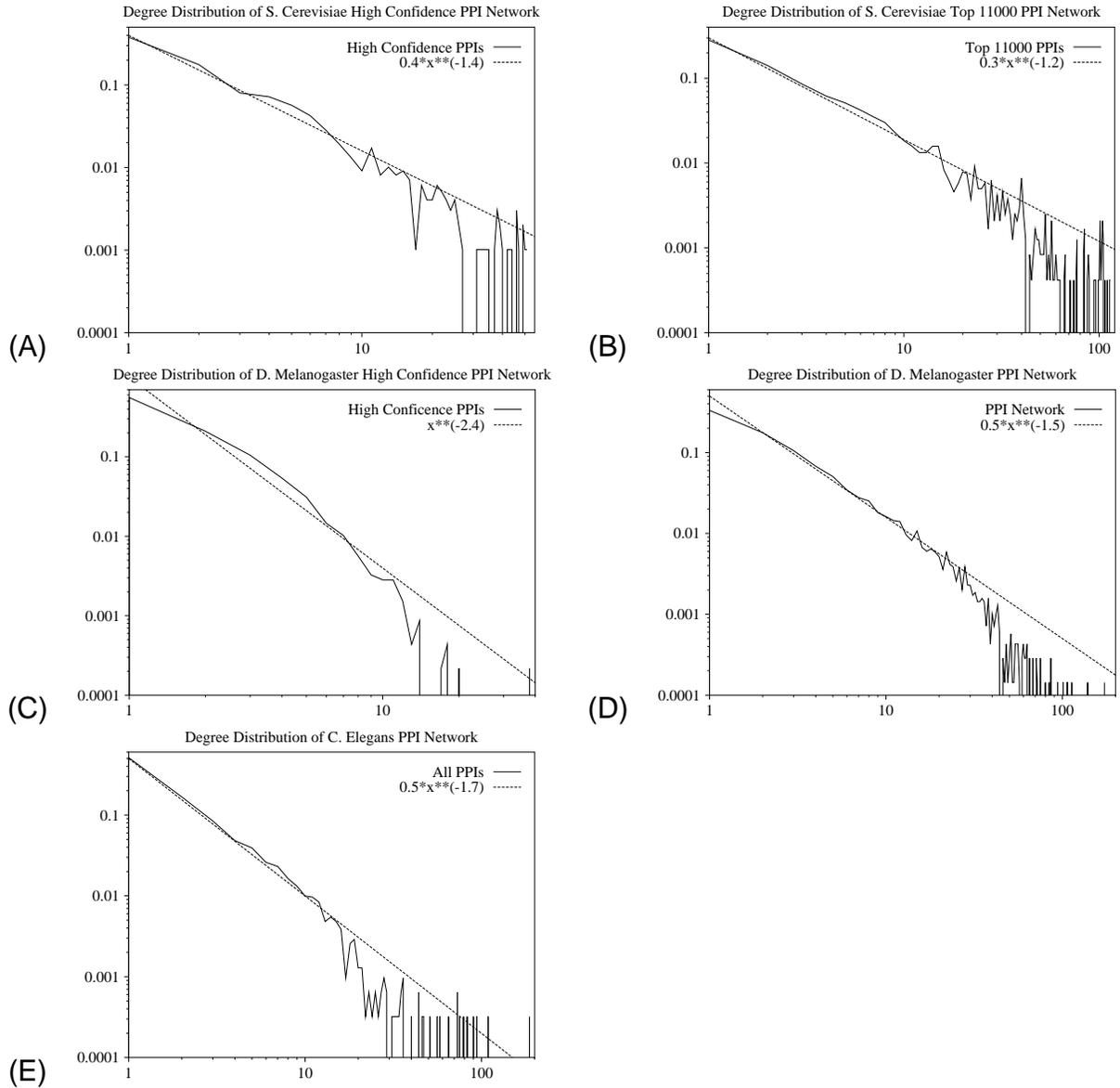

Figure 12: Degree distributions for PPI networks. The abscissa for each panel is node degree $x$ and the ordinate is the probability distribution of degrees, i.e., the fraction of nodes having degree $x$. **A.** High confidence *S. cerevisiae* PPI network [29]. **B.** *S. cerevisiae* "top 11000" PPI network [29]. **C.** *D. melanogaster* high confidence PPI network [27]. **D.** Entire currently available *D. melanogaster* PPI network [27]. **E.** *C. elegans* PPI network [59].

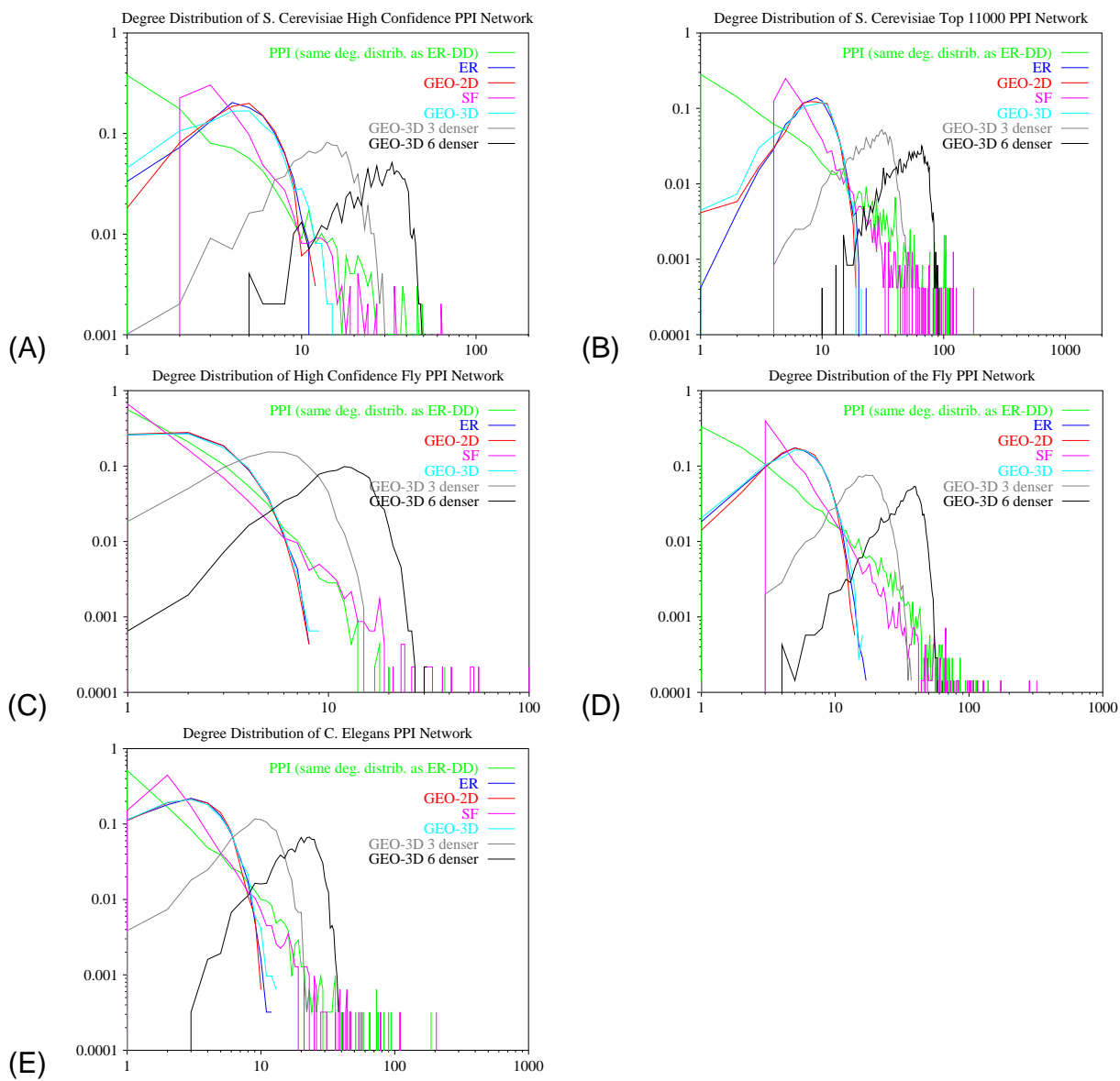

Figure 13: Degree distributions for PPI networks *versus* the degree distributions of the corresponding random graphs. Degree distribution of only one random graph belonging to each of the network models is drawn in each panel, since this is just an illustration. The abscissa for each panel is node degree $x$ and the ordinate is the probability distribution of degrees, i.e., the fraction of nodes having degree $x$. **A.** High confidence *S. cerevisiae* PPI network [29]. **B.** *S. cerevisiae* "top 11000" PPI network [29]. **C.** *D. melanogaster* high confidence PPI network[27]. **D.** Entire currently available *D. melanogaster* PPI network [27]. **E.** *C. elegans* PPI network [59].

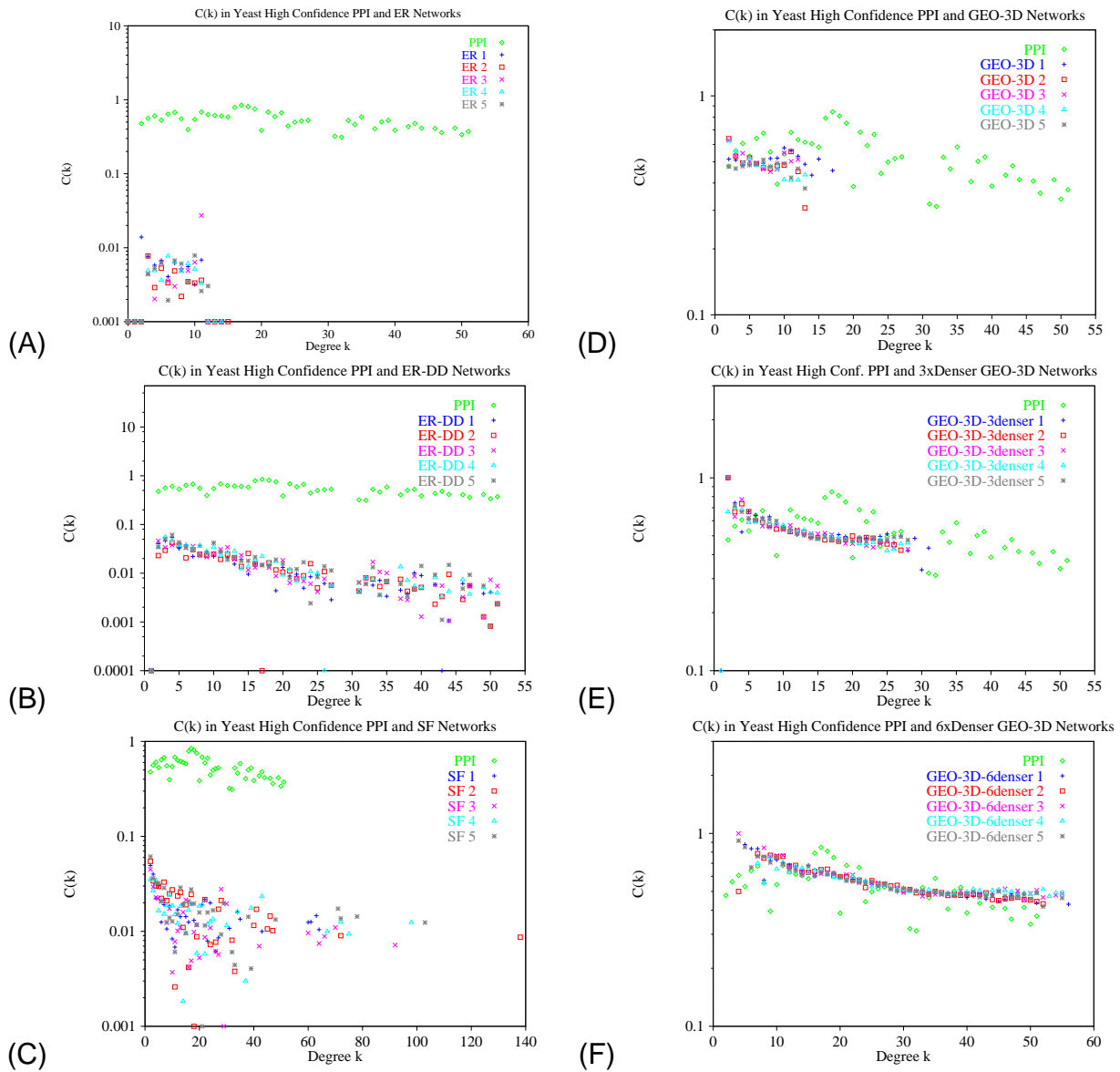

Figure 14: Comparison of degree $k$ node average clustering coefficients $C(k)$ as a function of degree $k$ in high confidence *S. cerevisiae* PPI network [29] and corresponding random graphs: **A.** PPI network *versus* five corresponding ER networks. **B.** PPI network *versus* five corresponding ER-DD networks. **C.** PPI network *versus* five corresponding SF networks. **D.** PPI network *versus* five corresponding GEO-3D networks. **E.** PPI network *versus* five corresponding GEO-3D networks which are 3 times denser than the PPI network. **F.** PPI network *versus* five corresponding GEO-3D networks which are 6 times denser than the PPI network.

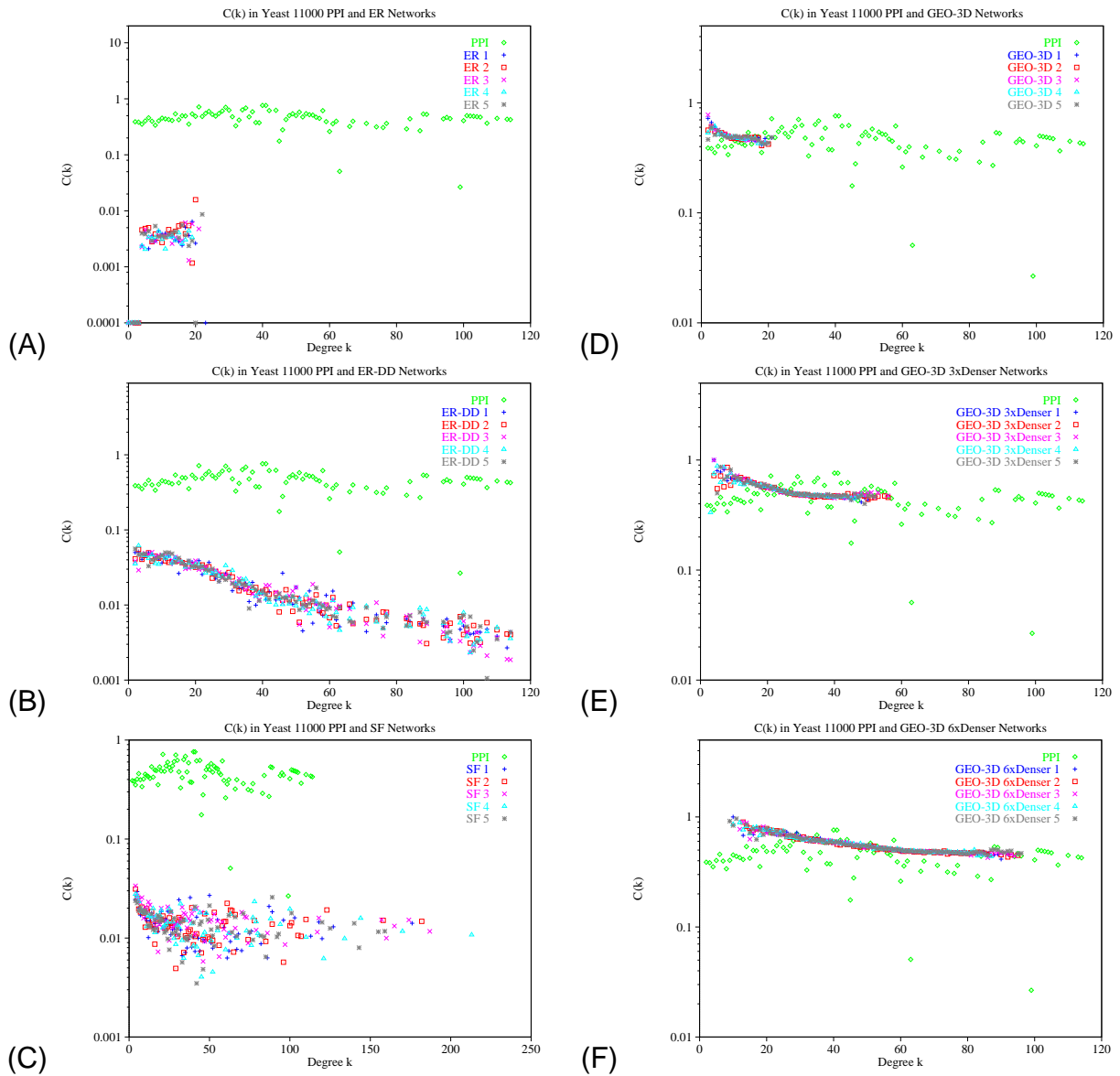

Figure 15: Comparison of degree $k$ node average clustering coefficients $C(k)$ as a function of degree $k$ in the "top 11000" *S. cerevisiae* PPI network [29] and corresponding random graphs: **A.** PPI network *versus* five corresponding ER networks. **B.** PPI network *versus* five corresponding ER-DD networks. **C.** PPI network *versus* five corresponding SF networks. **D.** PPI network *versus* five corresponding GEO-3D networks. **E.** PPI network *versus* five corresponding GEO-3D networks which are 3 times denser than the PPI network. **F.** PPI network *versus* five corresponding GEO-3D networks which are 6 times denser than the PPI network.

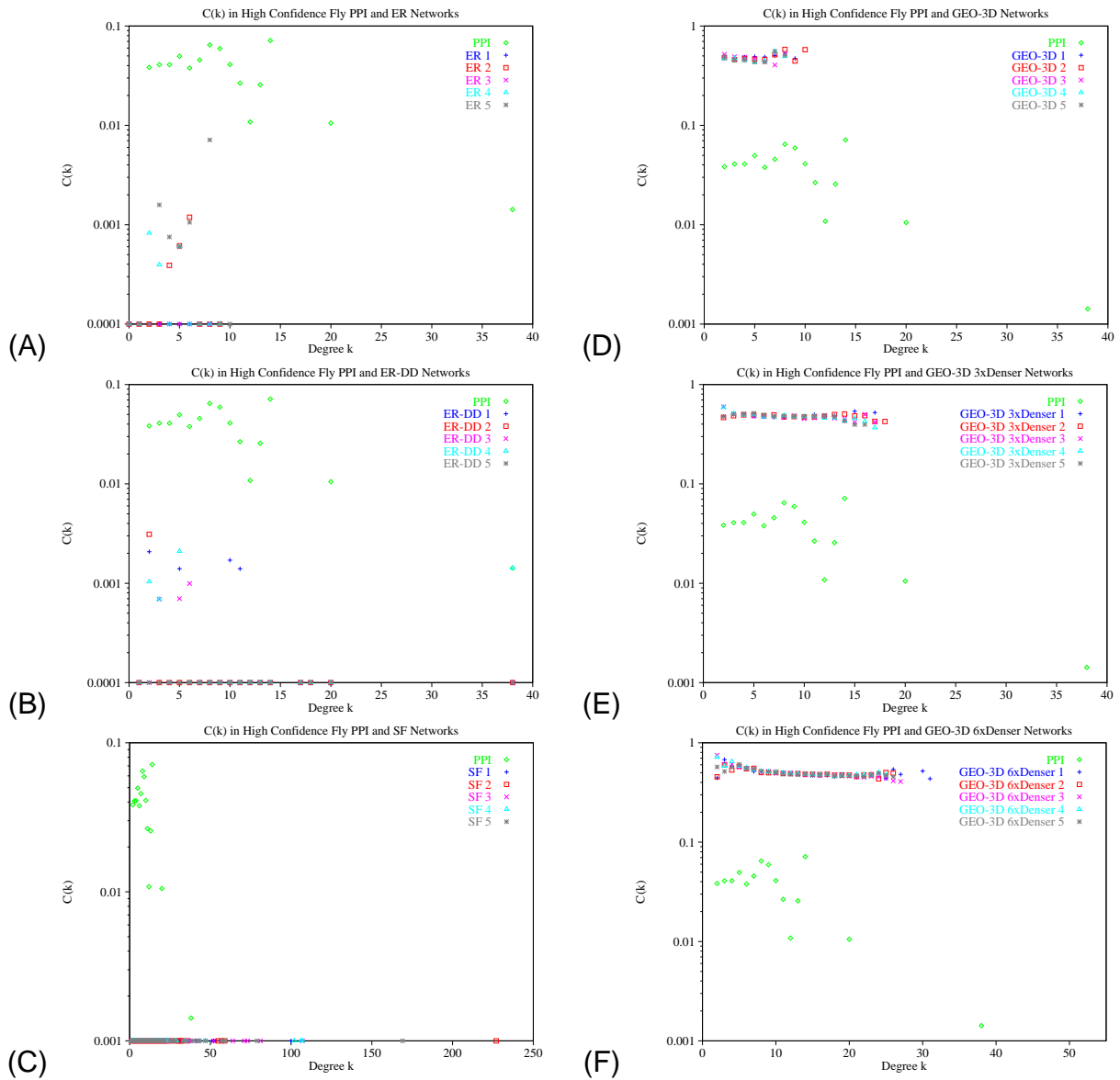

Figure 16: Comparison of degree $k$ node average clustering coefficients $C(k)$ as a function of degree $k$ in the high-confidence *D. melanogaster* PPI network [27] and corresponding random graphs: **A.** PPI network *versus* five corresponding ER networks. **B.** PPI network *versus* five corresponding ER-DD networks. **C.** PPI network *versus* five corresponding SF networks. **D.** PPI network *versus* five corresponding GEO-3D networks. **E.** PPI network *versus* five corresponding GEO-3D networks which are 3 times denser than the PPI network. **F.** PPI network *versus* five corresponding GEO-3D networks which are 6 times denser than the PPI network.

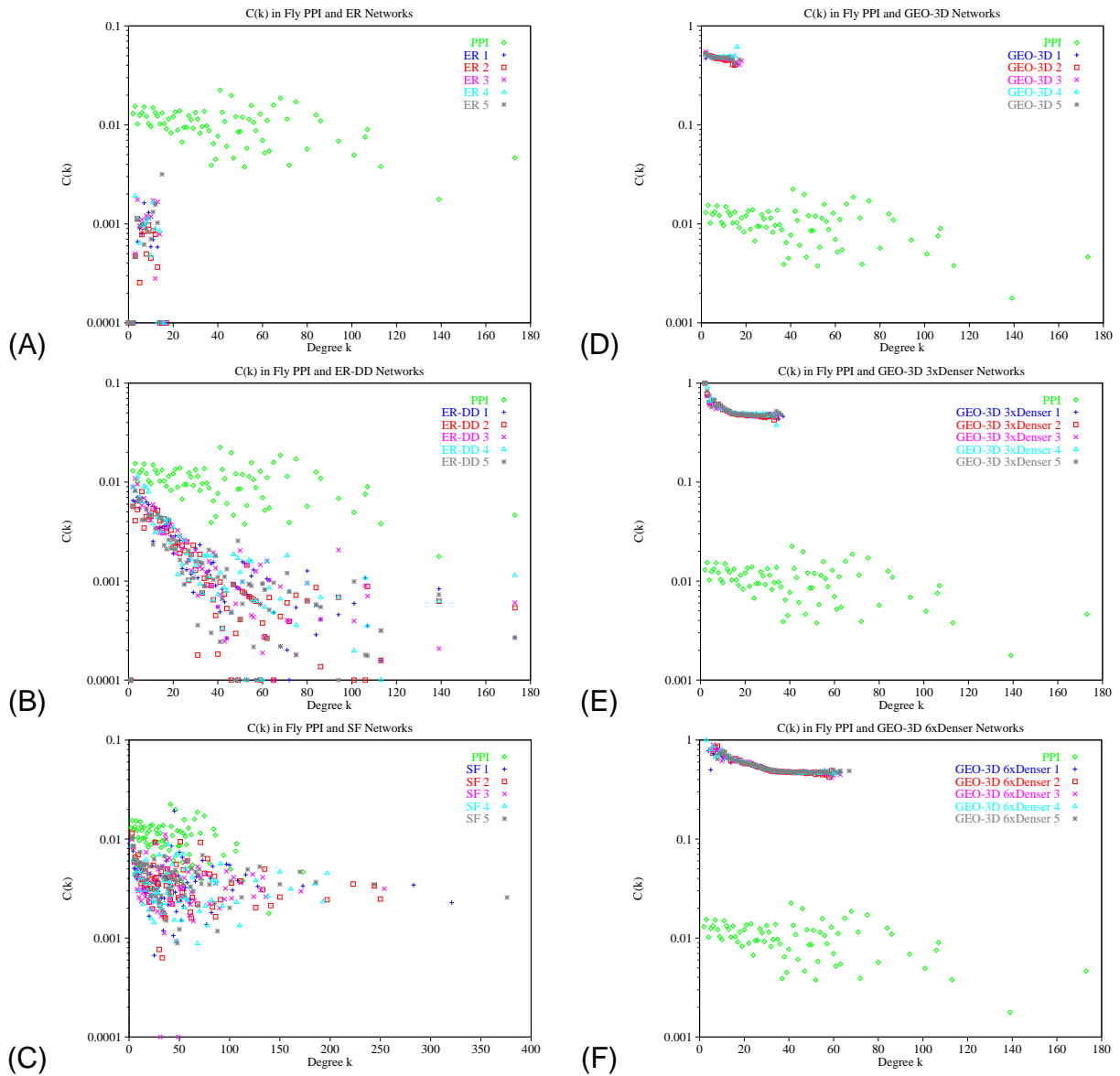

Figure 17: Comparison of degree $k$ node average clustering coefficients $C(k)$ as a function of degree $k$ in the entire currently available *D. melanogaster* PPI network [27] and corresponding random graphs: **A.** PPI network *versus* five corresponding ER networks. **B.** PPI network *versus* five corresponding ER-DD networks. **C.** PPI network *versus* five corresponding SF networks. **D.** PPI network *versus* five corresponding GEO-3D networks. **E.** PPI network *versus* five corresponding GEO-3D networks which are 3 times denser than the PPI network. **F.** PPI network *versus* five corresponding GEO-3D networks which are 6 times denser than the PPI network.

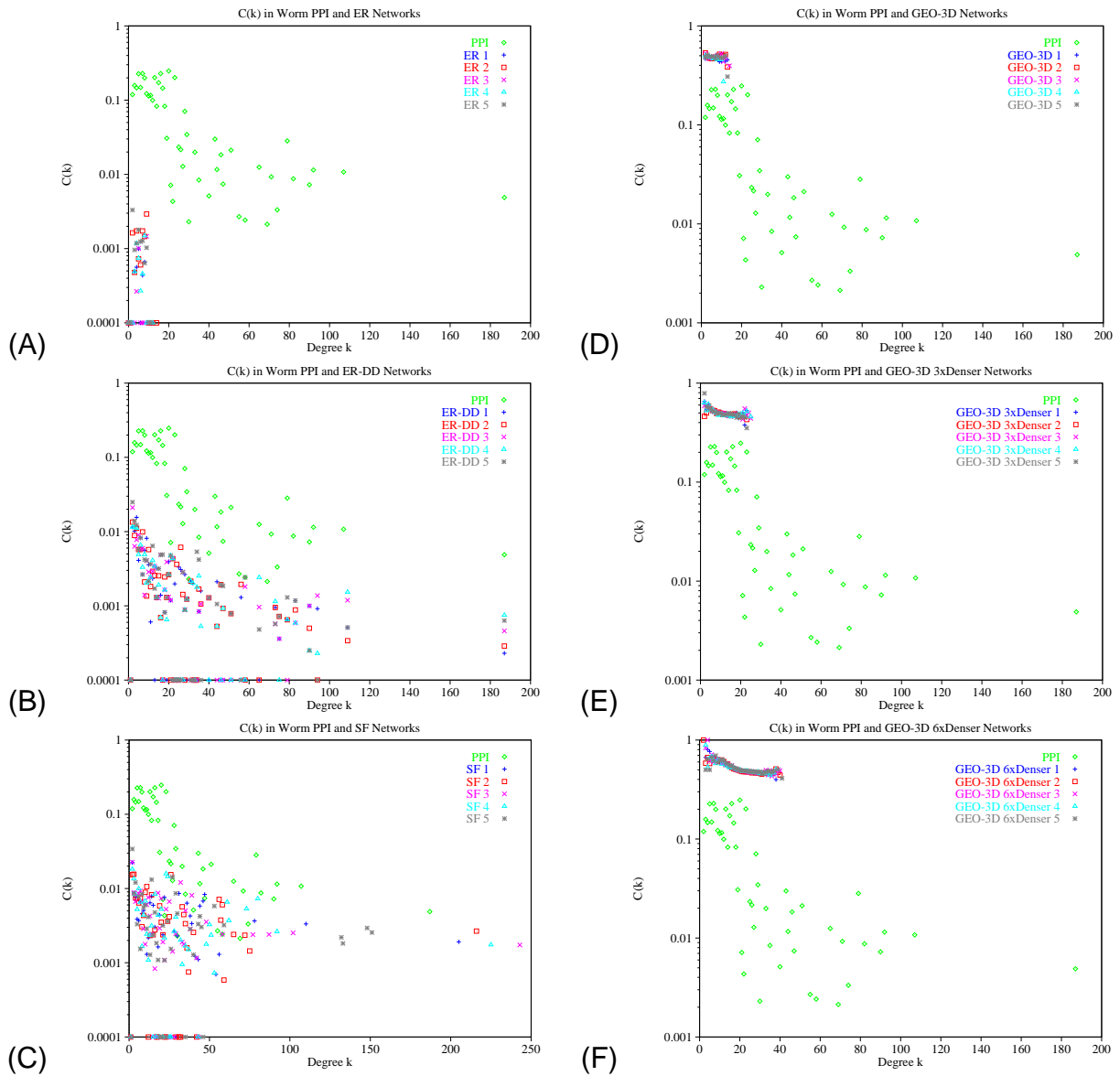

Figure 18: Comparison of degree $k$ node average clustering coefficients $C(k)$ as a function of degree $k$ in the *C. elegans* PPI network [59] and corresponding random graphs: **A.** PPI network *versus* five corresponding ER networks. **B.** PPI network *versus* five corresponding ER-DD networks. **C.** PPI network *versus* five corresponding SF networks. **D.** PPI network *versus* five corresponding GEO-3D networks. **E.** PPI network *versus* five corresponding GEO-3D networks which are 3 times denser than the PPI network. **F.** PPI network *versus* five corresponding GEO-3D networks which are 6 times denser than the PPI network.

# C Supplementary Tables

|            | Add      | Delete    | Rewire    |
|------------|----------|-----------|-----------|
| 10% graph 1 | 2.013974 | 10.027040 | 10.286293 |
| 10% graph 2 | 2.195439 | 10.722417 | 13.761059 |
| 10% graph 3 | 2.286288 | 9.030401  | 12.536496 |
| 10% graph 4 | 2.365492 | 11.882987 | 11.471204 |
| 10% graph 5 | 2.610209 | 10.050913 | 11.269877 |
| 20% graph 1 | 3.805339 | 17.021142 | 21.919662 |
| 20% graph 2 | 3.772996 | 16.022671 | 18.817653 |
| 20% graph 3 | 3.740578 | 17.997643 | 19.853645 |
| 20% graph 4 | 5.229768 | 16.026717 | 21.458437 |
| 20% graph 5 | 4.351935 | 19.315072 | 21.813793 |
| 30% graph 1 | 7.149713 | 24.846386 | 24.352380 |
| 30% graph 2 | 5.925923 | 20.917719 | 25.774724 |
| 30% graph 3 | 5.966197 | 19.724582 | 22.898066 |
| 30% graph 4 | 5.788158 | 25.183875 | 26.607413 |
| 30% graph 5 | 7.039671 | 24.201879 | 23.885723 |

Table 2: Graphlet frequency distances between high-confidence yeast PPI network and the perturbed networks with 10, 20, and 30 percent of edges added, deleted, or rewired at random. For example, 2.013974 in the top left-most field indicates that the first perturbed network with 10% of edges added at random to the yeast high-confidence PPI network is at distance 2.013974 from the yeast high-confidence PPI network.

| | Yeast High Conf. | Yeast top 11000 | Fruitfly High Conf. | Fruitfly Entire |
|---|---|---|---|---|
| ER 1 | 171.033026 | 213.492367 | 118.708767 | 109.821280 |
| ER 2 | 179.223899 | 207.739178 | 98.237245 | 112.953102 |
| ER 3 | 171.104348 | 208.926147 | 124.543077 | 111.693394 |
| ER 4 | 171.595828 | 217.690919 | 98.476123 | 100.810212 |
| ER 5 | 169.166791 | 213.818171 | 91.245012 | 100.810212 |
| ER-DD 1 | 164.333759 | 157.611121 | 107.402586 | 98.360966 |
| ER-DD 2 | 164.238223 | 159.114484 | 139.985358 | |
| ER-DD 3 | 162.804537 | 158.042483 | 112.846698 | |
| ER-DD 4 | 162.184038 | 161.413934 | 110.192676 | |
| ER-DD 5 | 154.193839 | 158.009708 | 133.969689 | |
| SF 1 | 142.916451 | 125.505065 | 229.710937 | 50.552265 |
| SF 2 | 142.560031 | | 297.435880 | |
| SF 3 | 130.977196 | | 244.786779 | |
| SF 4 | 125.978748 | | 252.437345 | |
| SF 5 | 125.468572 | | 270.795025 | |
| GEO-2D 1 | 35.466770 | 51.385464 | 89.897281 | 156.019444 |
| GEO-2D 2 | 38.960880 | 51.219103 | 90.872934 | 156.790793 |
| GEO-2D 3 | 36.780601 | 50.551726 | 89.040371 | 156.157409 |
| GEO-2D 4 | 37.542637 | 51.150659 | 92.284847 | 156.173940 |
| GEO-2D 5 | 37.014469 | 51.090161 | 93.675709 | 156.961472 |
| GEO-3D 1 | 27.250862 | 40.791194 | 90.392713 | 142.897851 |
| GEO-3D 2 | 33.471204 | 41.070521 | 92.647295 | 142.663669 |
| GEO-3D 3 | 34.264573 | 41.526624 | 87.959564 | 142.716685 |
| GEO-3D 4 | 31.707761 | 41.299082 | 86.764787 | |
| GEO-3D 5 | 31.399684 | 41.290330 | 90.067785 | |
| GEO-4D 1 | 32.338902 | 38.784356 | 80.767414 | 134.651342 |
| GEO-4D 2 | 30.963391 | 40.182807 | 82.510407 | 134.447379 |
| GEO-4D 3 | 30.738171 | 39.168425 | 82.170915 | 135.150400 |
| GEO-4D 4 | 31.914085 | 39.022422 | 75.953468 | |
| GEO-4D 5 | 31.426417 | 38.647461 | 82.780935 | |
| GEO-2D 3 denser 1 | 30.559877 | | | |
| GEO-3D 3 denser 1 | 18.381662 | | | |
| GEO-3D 3 denser 2 | 19.637330 | | | |
| GEO-3D 3 denser 3 | 20.169385 | | | |
| GEO-3D 3 denser 4 | 20.117049 | | | |
| GEO-3D 3 denser 5 | 19.464922 | | | |
| GEO-3D 6 denser 1 | 20.501105 | | | |
| GEO-3D 6 denser 2 | 20.130138 | | | |
| GEO-3D 6 denser 3 | 19.865880 | | | |
| GEO-3D 6 denser 4 | 20.106220 | | | |
| GEO-3D 6 denser 5 | 20.563879 | | | |

Table 3: Graphlet frequency distances between PPI networks and the corresponding random networks. Rows represent network types (for example, PPI, ER, ER-DD etc.) and columns represent the corresponding data set. Some values are missing due to the lack of computing power needed for exhaustive graphlet searches in large networks (the values of some ER-DD and SF model networks corresponding to the larger fruitfly and yeast PPI networks), or because they are not relevant for our analysis (the remaining missing values). The last 11 rows corresponds to 2- and 3-dimensional geometric random graphs with the same number of nodes, but approximately three ("3 denser") and six ("6 denser") times as many edges as the corresponding PPI graph.

|  | Yeast High Conf. | Yeast top 11000 | Fruitfly High Conf. | Fruitfly Entire | Worm Entire |
|---|---|---|---|---|---|
| PPI network | 5.193914 | 4.931596 | 9.436305 | 4.474159 | 4.961623 |
| ER 1 | 4.483198 | 3.757808 | 10.944241 | 5.273031 | 6.558142 |
| ER 2 | 4.459922 | 3.758895 | 11.235685 | 5.265893 | 6.568807 |
| ER 3 | 4.475276 | 3.758236 | 11.193257 | 5.267282 | 6.572835 |
| ER 4 | 4.472878 | 3.757333 | 11.094230 | 5.266617 | 6.536964 |
| ER 5 | 4.468909 | 3.754669 | 10.881638 | 5.269882 | 6.560432 |
| ER-DD 1 | 3.655439 | 3.339231 | 12.923071 | 4.232629 | 4.668186 |
| ER-DD 2 | 3.661022 | 3.326563 | 13.265843 | 4.232984 | 4.673229 |
| ER-DD 3 | 3.657886 | 3.336046 | 12.501059 | 4.234407 | 4.654749 |
| ER-DD 4 | 3.659162 | 3.334507 | 13.230065 | 4.230353 | 4.656498 |
| ER-DD 5 | 3.655491 | 3.330596 | 12.821002 | 4.229300 | 4.671363 |
| SF 1 | 3.758362 | 3.324137 | 9.333287 | 4.164283 | 4.844665 |
| SF 2 | 3.624404 | 3.340193 | 7.779545 | 4.127975 | 4.855700 |
| SF 3 | 3.682346 | 3.311710 | 8.359006 | 4.202195 | 4.756184 |
| SF 4 | 3.708943 | 3.304504 | 8.056372 | 4.191048 | 4.811073 |
| SF 5 | 3.659814 | 3.322626 | 8.866105 | 4.124067 | 4.789687 |
| GEO-2D 1 | 26.745117 | 20.363820 | 3.612093 | 49.954521 | 12.717975 |
| GEO-2D 2 | 23.556261 | 20.154389 | 3.517257 | 51.309693 | 18.533181 |
| GEO-2D 3 | 23.329942 | 20.375800 | 3.465955 | 51.263290 | 11.430562 |
| GEO-2D 4 | 22.565619 | 19.955981 | 4.012000 | 51.682084 | 17.959253 |
| GEO-2D 5 | 24.218951 | 19.940577 | 3.814337 | 52.261661 | 14.537689 |
| GEO-3D 1 | 11.891852 | 10.050567 | 6.123026 | 19.457399 | 27.521373 |
| GEO-3D 2 | 11.756614 | 10.215907 | 5.969310 | 19.379770 | 28.872684 |
| GEO-3D 3 | 10.987444 | 10.160251 | 4.836349 | 19.349904 | 30.852283 |
| GEO-3D 4 | 11.556084 | 10.041895 | 8.978468 | 19.061157 | 25.822047 |
| GEO-3D 5 | 12.429974 | 10.211669 | 5.596553 | 18.934693 | 32.476751 |
| GEO-4D 1 | 8.298618 | 7.370048 | 21.742505 | 12.491604 | 16.391512 |
| GEO-4D 2 | 8.034480 | 7.415314 | 17.325701 | 12.392516 | 16.166654 |
| GEO-4D 3 | 8.337146 | 7.351384 | 22.348251 | 12.457894 | 16.547705 |
| GEO-4D 4 | 8.263728 | 7.459610 | 9.684145 | 12.384512 | 16.261795 |
| GEO-4D 5 | 8.596852 | 7.457612 | 9.785776 | 12.659957 | 15.665861 |
| GEO-2D 3 denser 1 | 9.218131 | 9.916689 | 43.172889 | 21.735305 | 22.052473 |
| GEO-3D 3 denser 1 | 5.847585 | 5.783494 | 16.370559 | 10.411516 | 10.440371 |
| GEO-3D 3 denser 2 | 5.954403 | 5.847790 | 16.499478 | 10.336676 | 10.447413 |
| GEO-3D 3 denser 3 | 5.922108 | 5.830004 | 15.995224 | 10.344811 | 10.339636 |
| GEO-3D 3 denser 4 | 5.969824 | 5.823592 | 16.301939 | 10.397208 | 10.469520 |
| GEO-3D 3 denser 5 | 5.902647 | 5.882081 | 16.639121 | 10.293883 | 10.378750 |
| GEO-3D 6 denser 1 | 4.248278 | 4.448740 | 10.432986 | 7.541601 | 7.161603 |
| GEO-3D 6 denser 2 | 4.213818 | 4.414169 | 10.732926 | 7.531956 | 7.157405 |
| GEO-3D 6 denser 3 | 4.298536 | 4.412463 | 10.576001 | 7.576620 | 7.267117 |
| GEO-3D 6 denser 4 | 4.251699 | 4.414525 | 10.670042 | 7.544968 | 7.252137 |
| GEO-3D 6 denser 5 | 4.266613 | 4.393500 | 10.620819 | 7.577315 | 7.295091 |

Table 4: Diameters of PPI networks and their corresponding ER, ER-DD, SF, GEO-2D, GEO-3D, and GEO-4D random networks. Only distances between nodes in the same connected component are reported. Rows represent network types (for example, PPI, ER, ER-DD etc.) and columns represent the corresponding data set. "3 denser" and "6 denser" have the same meaning as in Supplementary Table 3.

|                  | Yeast High Conf. | Yeast top 11000 | Fruitfly High Conf. | Fruitfly Entire | Worm Entire |
|------------------|------------------|-----------------|---------------------|-----------------|-------------|
| PPI network      | 0.343733         | 0.304902        | 0.017923            | 0.008434        | 0.069764    |
| ER 1             | 0.006276         | 0.003471        | 0.000000            | 0.000820        | 0.000369    |
| ER 2             | 0.003828         | 0.003694        | 0.000072            | 0.000632        | 0.000990    |
| ER 3             | 0.003721         | 0.003283        | 0.000000            | 0.001034        | 0.000191    |
| ER 4             | 0.004622         | 0.003177        | 0.000311            | 0.000972        | 0.000477    |
| ER 5             | 0.004372         | 0.003795        | 0.000406            | 0.000807        | 0.001447    |
| ER-DD 1          | 0.021663         | 0.029159        | 0.000487            | 0.003655        | 0.003763    |
| ER-DD 2          | 0.016042         | 0.028207        | 0.000652            | 0.003210        | 0.005372    |
| ER-DD 3          | 0.020910         | 0.026906        | 0.000109            | 0.004831        | 0.003904    |
| ER-DD 4          | 0.021649         | 0.027990        | 0.000355            | 0.004874        | 0.004419    |
| ER-DD 5          | 0.022328         | 0.029896        | 0.000000            | 0.003544        | 0.004275    |
| SF 1             | 0.032902         | 0.020945        | 0.000000            | 0.007449        | 0.012701    |
| SF 2             | 0.035412         | 0.021071        | 0.000000            | 0.008452        | 0.011027    |
| SF 3             | 0.028359         | 0.024110        | 0.000000            | 0.007440        | 0.013002    |
| SF 4             | 0.026749         | 0.021944        | 0.000000            | 0.006447        | 0.012038    |
| SF 5             | 0.037393         | 0.019636        | 0.000000            | 0.007994        | 0.018236    |
| GEO-2D 1         | 0.561738         | 0.579556        | 0.323860            | 0.562504        | 0.469442    |
| GEO-2D 2         | 0.551851         | 0.578821        | 0.336352            | 0.560252        | 0.477439    |
| GEO-2D 3         | 0.542434         | 0.585639        | 0.329381            | 0.552721        | 0.474385    |
| GEO-2D 4         | 0.534709         | 0.581867        | 0.337864            | 0.555136        | 0.477508    |
| GEO-2D 5         | 0.529600         | 0.577737        | 0.332533            | 0.565060        | 0.474463    |
| GEO-3D 1         | 0.470745         | 0.498465        | 0.284394            | 0.469083        | 0.406792    |
| GEO-3D 2         | 0.482708         | 0.497322        | 0.280968            | 0.470272        | 0.416314    |
| GEO-3D 3         | 0.489211         | 0.500503        | 0.300616            | 0.473269        | 0.414913    |
| GEO-3D 4         | 0.487208         | 0.496027        | 0.258490            | 0.472467        | 0.416880    |
| GEO-3D 5         | 0.456253         | 0.496865        | 0.266211            | 0.470765        | 0.408034    |
| GEO-4D 1         | 0.394535         | 0.430832        | 0.233926            | 0.395276        | 0.352444    |
| GEO-4D 2         | 0.389667         | 0.428035        | 0.243421            | 0.400708        | 0.345016    |
| GEO-4D 3         | 0.406101         | 0.428948        | 0.230490            | 0.395749        | 0.360883    |
| GEO-4D 4         | 0.414717         | 0.442019        | 0.227551            | 0.393788        | 0.333224    |
| GEO-4D 5         | 0.406548         | 0.433822        | 0.217219            | 0.399770        | 0.354642    |
| GEO-2D 3 denser 1 | 0.608068        | 0.602326        | 0.563866            | 0.589434        | 0.584729    |
| GEO-3D 3 denser 1 | 0.523472        | 0.518105        | 0.470257            | 0.498982        | 0.500098    |
| GEO-3D 3 denser 2 | 0.520203        | 0.519829        | 0.472369            | 0.498337        | 0.499070    |
| GEO-3D 3 denser 3 | 0.522614        | 0.517125        | 0.473605            | 0.496640        | 0.502030    |
| GEO-3D 3 denser 4 | 0.520670        | 0.516553        | 0.478754            | 0.504280        | 0.493827    |
| GEO-3D 3 denser 5 | 0.517115        | 0.520450        | 0.479796            | 0.497585        | 0.500288    |
| GEO-3D 6 denser 1 | 0.538432        | 0.535339        | 0.492736            | 0.504445        | 0.506230    |
| GEO-3D 6 denser 2 | 0.540154        | 0.531674        | 0.496888            | 0.503071        | 0.505796    |
| GEO-3D 6 denser 3 | 0.543724        | 0.534059        | 0.495386            | 0.505574        | 0.508243    |
| GEO-3D 6 denser 4 | 0.537005        | 0.532969        | 0.500067            | 0.502805        | 0.506092    |
| GEO-3D 6 denser 5 | 0.539893        | 0.534842        | 0.493908            | 0.505338        | 0.514109    |

Table 5: Clustering coefficients of PPI networks and their corresponding ER, ER-DD, SF, GEO-2D, GEO-3D, and GEO-4D random networks. Rows represent network types (for example, PPI, ER, etc.) and columns represent the corresponding data set.

1. Erdös, P. and Rényi, A. On random graphs. *Publicationes Mathematicae* **6**, 290–297 (1959).

2. Erdös, P. and Rényi, A. On the evolution of random graphs. *Publ. Math. Inst. Hung. Acad. Sci.* **5**, 17–61 (1960).

3. Erdös, P. and Rényi, A. On the strength of connectedness of a random graph. *Acta Mathematica Scientia Hungary* **12**, 261–267 (1961).

4. Watts, D. J. and Strogatz, S. H. Collective dynamics of 'small-world' networks. *Nature* **393**, 440–442 (1998).

5. Barabási, A. L. and Albert, R. Emergence of scaling in random networks. *Science* **286**(5439), 509–12 (1999).

6. Ravasz, E., Somera, A. L., Mongru, D. A., Oltvai, Z. N., and Barabási, A.-L. Hierarchical organization of modularity in metabolic networks. *Science* **297**, 1551–5 (2002).

7. Newman, M. E. J. The structure and function of complex networks. *SIAM Review* **45**(2), 167–256 (2003).

8. Barabási, A.-L. and Oltvai, Z. N. Network biology: Understanding the cell's functional organization. *Nature Reviews* **5**, 101–113 (2004).

9. Albert, R. and A.-L., B. Statistical mechanics of complex networks. *Reviews of Modern Physics* **74**, 47–97 (2002).

10. Strogatz, S. H. Exploring complex networks. *Nature* **410**, 268–276 (2001).

11. Lappe, M. and Holm, L. Unraveling protein interaction networks with near-optimal efficiency. *Nature Biotechnology* **22**(1), 98–103 (2004).

12. West, D. B. *Introduction to Graph Theory*. Prentice Hall, Upper Saddle River, NJ., (1996).

13. Chung, F. and Lu, L. The average distances in random graphs with given expected degrees. *Proc. Natl. Acad. Sci. USA* **99**, 15879–15882 (2002).

14. Cohen, R. and Havlin, S. Scale-free networks are ultra small. *Physical Review Letters* **90**, 058701 (2003).

15. Ravasz, E. and Barabási, A.-L. Hierarchical organization in complex networks. *Phys. Rev. E Stat. Nonlin. Soft Matter Phys.* **67**, 026112 (2003).

16. Milo, R., Shen-Orr, S. S., Itzkovitz, S., Kashtan, N., Chklovskii, D., and Alon, U. Network motifs: simple building blocks of complex networks. *Science* **298**, 824–827 (2002).

17. Shen-Orr, S. S., Milo, R., Mangan, S., and Alon, U. Network motifs in the transcriptional regulation network of escherichia coli. *Nature Genetics* **31**, 64–68 (2002).

18. Itzkovitz, S., Milo, R., Kashtan, N., Ziv, G., and Alon, U. Subgraphs in random networks. *Physical Review E* **68**, 026127 (2003).

19. Milo, R., Itzkovitz, S., Kashtan, N., Levitt, R., Shen-Orr, S., Ayzenshtat, I., Sheffer, M., and Alon, U. Superfamilies of evolved and designed networks. *Science* **303**, 1538–1542 (2004).

20. Uetz, P., Giot, L., Cagney, G., Mansfield, T. A., Judson, R. S., Knight, J. R., Lockshon, D., Narayan, V., Srinivasan, M., Pochart, P., Qureshi-Emili, A., Li, Y., God-

win, B., Conover, D., Kalbfleish, T., Vijayadamodar, G., Yang, M., Johnston, M., Fields, S., and Rothberg, J. M. A comprehensive analysis of protein-protein interactions in saccharomyces cerevisiae. *Nature* **403**, 623–627 (2000).

21. Xenarios, I., Salwinski, L., Duan, X. J., Higney, P., Kim, S. M., and D., E. Dip: the database of interacting proteins. *Nucleic Acids Research* **28**(1), 289–291 (2000).

22. Ito, T., , Chiba, T., Ozawa, R., Yoshida, M., Hattori, M., and Sakaki, Y. A comprehensive two-hybrid analysis to explore the yeast protein interactome. *Proc Natl Acad Sci U S A* **98**(8), 4569–4574 (2001).

23. Jeong, H., Mason, S. P., Barabási, A. L., and Oltvai, Z. N. Lethality and centrality in protein networks. *Nature* **411**(6833), 41–2 (2001).

24. Maslov, S. and Sneppen, K. Specificity and stability in topology of protein networks. *Science* **296**(5569), 910–3 (2002).

25. Barabási, A.-L., Dezso, Z., Ravasz, E., Yook, Z.-H., and Oltvai, Z. N. Scale-free and hierarchical structures in complex networks. In *Sitges Proceedings on Complex Networks*, (2004). to appear.

26. Rain, J.-D., Selig, L., De Reuse, H., Battaglia, V., Reverdy, C., Simon, S., Lenzen, G., Petel, F., Wojcik, J., Schachter, V., Chemama, Y., Labigne, A., and Legrain, P. The protein-protein interaction map of helicobacter pylori. *Nature* **409**, 211–215 (2001).

27. Giot, L., Bader, J., Brouwer, C., Chaudhuri, A., Kuang, B., Li, Y., Hao, Y., Ooi, C., Godwin, B., Vitols, E., Vijayadamodar, G., Pochart, P., Machineni, H., Welsh, M., Kong, Y., Zerhusen, B., Malcolm, R., Varrone, Z., Collis, A., Minto, M., Burgess, S.,

McDaniel, L., Stimpson, E., Spriggs, F., Williams, J., Neurath, K., Ioime, N., Agee, M., Voss, E., Furtak, K., Renzulli, R., Aanensen, N., Carrolla, S., Bickelhaupt, E., Lazovatsky, Y., DaSilva, A., Zhong, J., Stanyon, C., Finley, R. J., White, K., Braverman, M., Jarvie, T., Gold, S., Leach, M., Knight, J., Shimkets, R., McKenna, M., Chant, J., and Rothberg, J. A protein interaction map of drosophila melanogaster. *Science* **302**(5651), 1727–1736 (2003).

28. Penrose, M. *Geometric Random Graphs*. Oxford Univeristy Press, (2003).

29. von Mering, C., Krause, R., Snel, B., Cornell, M., Oliver, S. G., Fields, S., and Bork, P. Comparative assessment of large-scale data sets of protein-protein interactions. *Nature* **417**(6887), 399–403 (2002).

30. Bettstetter, C. On the minimum node degree and connectivity of a wireless multi-hop network. In *Proceedings of the 3rd ACM international symposium on mobile ad hoc networking and computing*, 80–01, (2002).

31. Bollobas, B. *Random Graphs*. Academic, London, (1985).

32. Bender, E. A. and Canfield, E. R. The asymptotic number of labeled graphs with given degree sequences. *Journal of Combinatorial Theory A* **24**, 296–307 (1978).

33. Newman, M. E. J. Random graphs as models of networks. In Handbook of Graphs and Networks, Bornholdt, S. and Schuster, H. G., editors. Wiley-VHC, Berlin (2002).

34. Luczak, T. Component behavior near the critical point of the random graph process. *Random Structures and Algorithms* **1**, 287 (1990).

35. Molloy, M. and Reed, B. A critical point of random graphs with a given degree sequence. *Random Structures and Algorithms* **6**, 161–180 (1995).

36. Molloy, M. and Reed, B. The size of the largest component of a random graph on a fixed degree sequence. *Combinatorics, Probability and Computing* **7**, 295–306 (1998).

37. Aiello, W., Chung, F., and Lu, L. A random graph model for power law graphs. *Experimental Mathematics* **10**, 53–66 (2001).

38. Newman, M. E. J., Strogatz, S. H., and Watts, D. J. Random graphs with arbitrary degree distributions and their applications. *Physical Review E* **64**, 026118–1 (2001).

39. Wilf, H. S. *Generating Functionology*. Academic, Boston, (1990).

40. Przulj, N. Graph theory approaches to protein interaction data analysis. In Knowledge Discovery in High-Throughput Biological Domains, Jurisica, I. and Wigle, D., editors. (2004). to appear.

41. Barabási, A.-L., Albert, R., and Jeong, H. Mean-field theory for scale-free random networks. *Physica A* **272**, 173–197 (1999).

42. Faloutsos, M., Faloutsos, P., and Faloutsos, C. On power-law relationships of the internet topology. *Computer Communications Review* **29**, 251–262 (1999).

43. Jeong, H., Tombor, B., Albert, R., Oltvai, Z. N., and Barabási, A. L. The large-scale organization of metabolic networks. *Nature* **407**(6804), 651–4 (2000).

44. Abello, J., Buchsbaum, A., and Westbrook, J. A functional approach to external graph algorithms. *Lecture Notes in Computer Science* **1461**, 332–343 (1998).

45. Broder, A., Kumar, R., Maghoul, F., Raghavan, P., Rajagopalan, S., Stata, R., Tomkins, A., and Wiener, J. Graph structure of the web. *Computer Networks* **33**, 309–320 (2000).

46. Bollobas, B. and Riordan, O. The diameter of a scale-free random graph. *Combinatorica* (2001). to appear.

47. Krapivsky, P. L. and Redner, S. Organization of growing random networks. *Physical Review E* **63**, 066123–1 (2001).

48. Albert, R. and Barabási, A. L. Topology of evolving networks: local events and universality. *Phys Rev Lett* **85**(24), 5234–7 (2000).

49. Dorogovtsev, S. N. and Mendes, J. F. F. Evolution of networks with aging of sites. *Physical Review E* **62**, 1842–1845 (2000).

50. Krapivsky, P. L., Redner, S., and Leyvraz, F. Connectivity of growing random networks. *Physical Review Letters* **85**, 4629–4632 (2000).

51. Albert, R., Jeong, H., and Barabási, A.-L. Error and attack tolerance of complex networks. *Nature* **406**, 378–382 (2000).

52. Cohen, R., Erez, K., ben Avraham, D., and Havlin, S. Resilience of the internet to random breakdowns. *Physical Review Letters* **85**, 4626–4628 (2000).

53. Callaway, D. S., Newman, M. E. J., Strogatz, S. H., and Watts, D. J. Network robustness and fragility: percolation on random graphs. *Physical Review Letters* **85**, 5468–5471 (2000).

54. Bornholdt, S. and Ebel, H. World-wide web scaling exponent from simon's 1955 model. *Physical Review E* **64**, 046401 (2001).

55. Wagner, A. and Fell, D. The small world inside large metabolic networks. *Proc. Roy. Soc. London Series B* **268**, 1803–1810 (2001).

56. Diaz, J., Penrose, M. D., Petit, J., and Serna, M. Convergence theorems for some layout measures on random lattice and random geometric graphs. *Combinatorics, Probability and Computing* **10**, 489–511 (2000).

57. Diaz, J., Penrose, M. D., Petit, J., and Serna, M. Linear orderings of random geometric graphs. In *Workshop on Graph-Theoretic Concepts in Computer Science*, (1997).

58. Mehlhorn, K. and Naher, S. *Leda: A platform for combinatorial and geometric computing*. Cambridge University Press, (1999).

59. Li, S., Armstrong, C., Bertin, N., Ge, H., Milstein, S., Boxem, M., Vidalain, P.-O., Han, J.-D., Chesneau, A., Hao, T., Goldberg, DS Li, N., Martinez, M., Rual, J.-F., Lamesch, P., Xu, L., Tewari, M., Wong, S., Zhang, L., Berriz, G., Jacotot, L., Vaglio, P., Reboul, J., Hirozane-Kishikawa, T., Li, Q., Gabel, H., Elewa, A., Baumgartner, B., Rose, D., Yu, H., Bosak, S., Sequerra, R., Fraser, A., Mango, S., Saxton, W., Strome, S., van den Heuvel, S., Piano, F., Vandenhaute, J., Sardet, C., Gerstein, M., Doucette-Stamm, L., Gunsalus, K., Harper, J., Cusick, M., Roth, F., Hill, D., and Vidal, M. A map of the interactome network of the metazoan c. elegans. *Science* **303**, 540–543 (2004).